\documentclass{aa}  
\usepackage{natbib}
\usepackage[colorlinks=true, linkcolor=blue, urlcolor=blue, citecolor=blue, pdfborder={0 0 0}]{hyperref}
\bibpunct{(}{)}{;}{a}{}{,} 
\usepackage{tipa} 
\usepackage{graphicx}
\usepackage{txfonts}
\usepackage{lipsum}
\usepackage{caption}
\usepackage{subcaption}         
\usepackage{lscape}             
\usepackage{placeins}           
\usepackage{orcidlink} 
\usepackage{breqn} 

\begin{document}

 \title{A model of magnetised and rotating convection  \\ for stellar and planetary interiors}
\subtitle{}
 \author{L. Bessila \orcidlink{0009-0007-8721-7657} \inst{1}
          \and S. Mathis\inst{1}
          }

   \institute{Université Paris-Saclay, Université Paris Cité, CEA, CNRS, AIM, Gif-sur-Yvette, F-91191, France
  }

   \date{Received XX, Accepted YY}

  \abstract
   {Convection is a fundamental mechanism for energy transport in stars and planets, playing a pivotal role in shaping their structures and evolution. The Mixing-Length Theory, a monomodal approach to convection, is widely adopted and implemented in 1D stellar structure and evolution codes. However, it overlooks the combined effects of rotation and magnetic fields, which are ubiquitous across a wide range of stars and planets.}
   {To address this limitation, we extend the Mixing-Length Theory including both rotation and magnetic fields within a Cartesian set-up.}
   {Building on the work by \cite{stevenson_turbulent_1979}, we use a heat-flux maximisation principle, which amounts to selecting the convective mode that carries the most heat.}
   {Our findings show that both rotation and magnetic fields individually tend to suppress convection. However, when combined, they can enhance convection strength under certain conditions. We derive expressions for the root-mean-square (rms) velocity, characteristic length scale, and degree of superadiabaticity as functions of the rotation rate and magnetic field strength. These results offer new insights for more accurately modeling convection and its impact on stellar and planetary structures in one-dimensional and forthcoming multi-dimensional evolution models.}
   {}

   \maketitle


\everymath{\displaystyle}

\def\v{{\boldsymbol{v}}}
\def\u{{\boldsymbol{u}}}
\def\uosc{{\boldsymbol{u}_{\rm osc}}}
\def\ut{{\boldsymbol{U_{\rm t}}}}
\def \Bt{{\boldsymbol{B_t}}}
\def \B{{\boldsymbol{B}}}
\def \Bo{{\boldsymbol{B_0}}}
\def \bosc{{\boldsymbol{b_{\rm osc}}}}
\def \Om{{\boldsymbol{\Omega}}}

\def\et{{\boldsymbol{e_{\theta}}}}
\def \er{{\boldsymbol{e_{r}}}}
\def\ep{{\boldsymbol{e_{\varphi}}}}
\def\ex{{\boldsymbol{e_{x}}}}
\def \ey{{\boldsymbol{e_{y}}}}
\def\ez{{\boldsymbol{e_{z}}}}
\def \di{\boldsymbol{\nabla} \cdot}
\def \nab{\boldsymbol{\nabla}}
\def \rot{\boldsymbol{\nabla} \times}
\newcommand{\adv}[2]{\ensuremath \left(\boldsymbol{#1}\cdot \boldsymbol{\nabla}\right) #2}

\newcommand{\Dd}[2]{\ensuremath\frac{\partial #1}{\partial#2}}
\newcommand{\Dt}[1]{\ensuremath\frac{\partial #1}{\partial t}}
\newcommand{\Dtt}[1]{\ensuremath\frac{\partial^2 #1}{\partial t^2}}
\newcommand{\vc}[1]{\ensuremath\boldsymbol{#1}}


\def \rhoo{(\rho_0 + \rho_t)}
\def \Bot{\B}



\section{Introduction}
\subsection{Modelling convection in stars and planets}
 All stars and many planets exhibit convective processes within their interiors during various evolutionary stages \citep[e.g.][]{busse_simple_1976, kippenhahn_stellar_2012, guervilly_convective_2019}. In regions where radiative energy transport becomes inefficient, convection emerges as the primary mechanism for heat transfer \citep[e.g.][]{miesch_turbulence_2009, kupka_modelling_2017, garaud_double-diffusive_2018}. Moreover, convection significantly influences the internal dynamics of these celestial bodies, facilitating the mixing of chemicals,  the generation of magnetic fields through dynamo action \citep[e.g.][]{brun_magnetism_2017, brun_differential_2017, brun_powering_2022} and playing a pivotal role in the redistribution of angular momentum \citep[see e.g.][]{brun_magnetism_2017}. An accurate treatment of convective zones is fundamental to modelling the structure and evolution of stars and planets at every stage of their lifetimes. However, convection involves turbulent flows spanning a vast range of spatial and temporal scales. It thus poses significant challenges to solving the underlying equations on evolutionary timescales \citep{kupka_modelling_2017}. Two main strategies exist to address these difficulties. On the one hand, one can simulate a specific interval in the life of a star or planet to model the detailed dynamics, albeit over limited time-scales due to computational constraints. These simulations can either cover a full spherical domain focusing on large-scale flows \citep[see e.g.][]{aubert_spherical_2017, aurnou_connections_2020, hotta_solar_2017, brun_magnetism_2017}, or a smaller interior portion, allowing for finer resolution of turbulent processes \citep[e.g.][]{julien_strongly_1998, barker_theory_2014, currie_convection_2020}. Such simulations help shed light on the complex dynamics of convection influenced by rotation and/or magnetism. On the other hand, parameterized semi-analytical models like the widely used Mixing-Length Theory \citep[hereafter MLT, e.g.][]{bohm-vitense_uber_1958, gough_calibration_1976, kippenhahn_stellar_2012} are of great use. This theory models convection as parcels of fluids, that travel a specified length before mixing with their surrounding environment. MLT provides a simplified formula for the temperature gradient needed to transport energy by convection, along with the resulting characteristic convective length scale and velocity. Despite its simplicity, MLT has provided interesting results in modelling the structure of stars and gaseous planets \citep[e.g.][]{baraffe_new_2015}. It is now implemented in a majority of stellar structure and evolution codes such as the Modules for Experiments in Stellar Astrophysics MESA \citep{paxton_modules_2011, paxton_modules_2013, paxton_modules_2015, paxton_modules_2018, paxton_modules_2019, jermyn_modules_2023}, the Dartmouth stellar evolution code \citep{dotter_dartmouth_2008}, the Geneva code \citep{eggenberger_geneva_2008}, the Starevol code \citep{palacios_rotational_2003, amard_first_2019}.
 This theory however suffers from well-known limitations. For example, its standard version fails to model the overshooting between convective and radiative zones and is not time-dependent \citep[see e.g.][for more details about the limitations of MLT]{renzini_embarrassments_1987, arnett_3d_2019}. Several authors have developed alternative approaches to modelling stellar convection to provide a more realistic physical picture. For instance, \cite{canuto_turbulent_1996} derived a Reynolds stresses averages model \citep[see also][]{xiong_nonlocal_1997, canuto_stellar_2011}. Another strategy is to extend the standard MLT including some additional physical processes. \cite{lesaffre_two-dimensional_2013} generalised MLT for axisymmetric 2D stellar evolution models, while \cite{jermyn_turbulence_2018} developed a turbulence closure based on MLT with 3D fluctuations against a 2D background. Both studies treat each convective mode as growing with its linear growth rate, before saturating at an amplitude set by the turbulent cascade. However, the standard MLT neglects the influence of rotation and magnetic fields.
 
 \subsection{The impact of rotation and magnetic fields on convection}
 \label{sec:rotmag}
Rotation influences convection in several ways: first, the onset of convection is modified in the presence of rotation \citep[e.g.][]{chandrasekhar_hydrodynamic_1961, eltayeb_hydromagnetic_1977}. In the linear stability analysis, convection starts whenever the Rayleigh number (which is the ratio of buoyancy over viscosity and heat diffusion) is above the critical Rayleigh number. The critical Rayleigh number increases with the rotation rate $\Omega$, which results in a stabilising effect: convection is more difficult to start in a rotating case \citep{chandrasekhar_hydrodynamic_1961}. These are not the only effects of rotation on convection: it reduces horizontal lengthscales \citep{stevenson_turbulent_1979, julien_statistical_2012, barker_theory_2014} and breaks the spherical symmetry, with motion aligned along the rotation axis following the Taylor-Proudman constraint \citep{proudman_motion_1916, taylor_motion_1917}. Other complex aspects of rotating convection, such as the establishment of zonal flows, have been widely explored in numerical simulations \citep[e.g.][]{julien_strongly_1998, julien_plumes_1999, julien_heat_2012, julien_nonlinear_2016, gastine_scaling_2016, aurnou_rotating_2015, aurnou_connections_2020}. Moreover, experimental work has been carried out in rotating tanks to provide scaling relations for the heat transport in rotating convection, such as \cite{aurnou_experiments_2001} and \cite{hadjerci_rapidly_2024}. 
From a theoretical point of view, a reformulation of MLT for the rapidly rotating case has been developed by \cite{stevenson_turbulent_1979} and \cite{flasar_turbulent_1978}, followed by \cite{augustson_model_2019} who included the diffusive effects in the previous modellings. This theoretical model is based on the assumption that the nonlinear state is dominated by the mode that transports the most heat, following \cite{malkus_heat_1954}. This description provides the convective velocity, convective length scale and degree of superabadiaticity set by the dominant linear mode. Various studies \citep{barker_theory_2014, currie_convection_2020, devries_tidal_2023} conducted non-linear direct numerical simulations of rotating convection in a Cartesian geometry: despite their simplicity, \cite{stevenson_turbulent_1979} prescriptions of Rotating Mixing-Length Theory (R-MLT) agree well with these simulations in overall. 
The R-MLT model from \cite{stevenson_turbulent_1979} or \cite{augustson_model_2019} has then been useful to understand the convective overshoot in 3D spherical rotating numerical simulations \citep{korre_dynamics_2021}, light-elements mixing in late-type stars \citep{dumont_lithium_2021}, convective core boundary in early-type stars \citep{michielsen_probing_2019}, as well as gravito-inertial modes excitation by turbulent convection \citep{augustson_model_2020}. Finally, it has been implemented in the MESA stellar evolution code \citep{ireland_radius_2018} to show that taking into account rotation in convection models can impact the overall structure and evolution of fully convective stars. When it comes to giant gaseous planets, \cite{fuentes_rotation_2023} showed that rotation reduces convective mixing, with a good agreement between their numerical simulations and \cite{stevenson_turbulent_1979} analytical prescriptions. The authors suggest that planetary rotation is another factor contributing to the longevity of primordial composition gradients in Jupiter, as observed by the Juno spacecraft.

When it comes to magnetic fields, they are also known to inhibit convection via the Lorentz force. Following the mechanical movement of a fluid parcel \citep{gough_influence_1966}, followed by \cite{mullan_are_2001} showed that the criterion for the convective instability is modified in the presence of a magnetic field, making convection more difficult to start. Another stability criterion for convection is based on linear stability analysis through the critical Rayleigh number: in the absence of rotation, the critical Rayleigh number with a magnetic field is higher than in a non-magnetised situation. Magnetic fields are then stabilising for the convective instability \citep[see e.g.][]{chandrasekhar_hydrodynamic_1961, eltayeb_hydromagnetic_1977, horn_elbert_2022}. One significant difference between magnetoconvection and rotating convection is that the magnetic fields provide an additional source of dissipation, while rotation induces vorticity and breaks horizontal symmetry \citep{ecke_hopf_1992, potherat_why_2014}. Using again the heat-flux maximisation principle from \cite{malkus_heat_1954}, \cite{stevenson_turbulent_1979} developed a Magnetic Mixing-Length Theory (M-MLT) approach in the asymptotic case of a high magnetic field. Unlike its rotating counterpart, this extension of MLT has not been directly confronted with numerical simulations to our knowledge. \cite{hotta_breaking_2018} showed that the root-mean-square (r.m.s.) velocity in convection is diminished in the presence of magnetic fields, which agrees qualitatively with the result of \cite{stevenson_turbulent_1979}. A complementary approach used in stellar physics is based on the stability criterion from \cite{gough_influence_1966}: \cite{mullan_are_2001} studied the effects of magnetism in 1D stellar models, including the effects of magnetic pressure in the adiabatic temperature gradient $\nabla_{\rm ad}$. With this method, several authors showed that magnetic fields can significantly affect the radius of low-mass stars by inhibiting convection or through magnetic spot coverage \citep[see e.g.][]{mullan_are_2001, chabrier_evolution_2007, macdonald_surface_2014, macdonald_magnetic_2019}. More recently, \cite{ginzburg_younger_2024} also implemented this \cite{gough_influence_1966} criterion to account for the effect of magnetic fields on the cooling of white dwarfs. We stress that the main difference between the stability criterion method from \cite{gough_influence_1966} and the M-MLT method from \cite{stevenson_turbulent_1979} is that the former is an "on-off" approach, while the latter accounts for the progressive diminution in the convection strength as a function of the amplitude of the magnetic field, in a region where convection is unstable \citep[see e.g. the discussion in][ Sec. 4]{bessila_stochastic_2024}.

\par However, when rotation and magnetic field act together, they have conflicting tendencies and can strengthen convection when the Lorentz force and the Coriolis acceleration have the same order of magnitude. When both rotation and magnetic fields are present, the linear stability of convection is more complex. In some cases, the critical Rayleigh number with \textit{both} rotation and magnetic fields can be lower than with either rotation \textit{or} magnetism \citep{chandrasekhar_hydrodynamic_1961, eltayeb_hydromagnetic_1977, horn_elbert_2022}, resulting in enhanced convective flows.

 \par In this study, we extend the previous work by \cite{stevenson_turbulent_1979} and \cite{augustson_model_2019} to explore the effects of rotation and magnetic fields on MLT through a theoretical analysis based on the heat-flux maximisation principle. We present a generalised heuristic model for convection in rotating and magnetised configurations. The foundational equations and principles, including the linearised Boussinesq equations of motion and heat-flux maximisation principle, are detailed in Section \ref{sec:model}. In Section \ref{sec:rmlt}, we detail the case of Rotating MLT and recover the previous results from \cite{stevenson_turbulent_1979} and \cite{augustson_model_2019}. In Section \ref{sec:mmlt}, we address the case of Magnetised MLT, bridging the gap between the low magnetic field and high magnetic field cases from \cite{stevenson_turbulent_1979}. Finally, in Section \ref{sec:rmmlt}, we present a model of MLT with both rotation and magnetic field. The paper concludes with a summary of findings and future research directions in Section \ref{sec:conclusion}.

\section{Heat-flux maximised convection model}
\label{sec:model}
As a first step, we adopt a local model in Cartesian coordinates, which amounts to treating convection as a localised dynamical process at a specific radius and latitude within a star or planet.

\subsection{Linear Boussinesq equations}
We consider an infinite layer of fluid with a vertical rotation $\boldsymbol{\Omega}$ and a vertical imposed magnetic field $\boldsymbol{B}$. Imposing a fixed constant magnetic field amounts to assuming that its temporal variations occur on timescales much longer than the turnover time of the convective motions. We use the Boussinesq approximation for a nearly incompressible fluid, which consists of neglecting density fluctuations except where multiplied by gravity in the buoyancy. This approximation is justified for convective motion with a much smaller characteristic length scale than density variations. The fluid has a small thermal expansion coefficient $\alpha = - \partial \ln \rho / \partial T \lvert_P$, where $\rho$ is the density, $T$ the temperature and $P$ the pressure. The fluid is confined between two infinite impenetrable plates that have different temperatures and are separated by a distance $\ell_0$. The diffusion coefficients in stars are relatively low, and \cite{augustson_model_2019} have shown that adding viscosity and thermal diffusion does not significantly change the result for R-MLT. Thus, we neglect as a first step the thermal diffusivity, viscous diffusivity and magnetic diffusivity. In this framework, we use the linearised equations of magnetohydrodynamics, omitting all non-linear terms in fluctuating quantities following \cite{stevenson_turbulent_1979}. The linearised Navier-Stokes equation within the Boussinesq approximation with rotation and an imposed magnetic field $\boldsymbol{B}$ is \citep[e.g.][]{chandrasekhar_hydrodynamic_1961}: 
\begin{equation}
    \frac{\partial \v}{\partial t}+ \v \cdot \nabla \v + 2 \Om \times \v = \frac{-1}{\rho} \nabla P - \boldsymbol{g} \alpha \theta + \frac{(\nabla \times \boldsymbol{b}) \times \B}{\mu_0 \rho}, 
    \label{eq:navier-lin}
\end{equation}
where $\v$ is the fluid velocity in the rotating frame, $p$, $\theta$ and $\boldsymbol{b}$ are the changes in pressure, temperature and magnetic field caused by the flow, respectively.
The linearised thermal diffusion equation is \citep[e.g.][]{chandrasekhar_hydrodynamic_1961}: 
\begin{equation}
    \frac{\partial \theta}{\partial t} = \beta v_z,
\end{equation}
where $\beta$ is the average deviation of the thermal gradient from the adiabatic value: 
\begin{equation}
    \beta = \frac{d T}{dz} + \frac{c_p}{g},
\end{equation}
where $T$ is the mean temperature and $c_p$ is the specific heat capacity at constant pressure.
In addition, the velocity field in the Boussinesq approximation is solenoidal: 
\begin{equation}
  \nabla \cdot \v = 0.  
\end{equation}
Finally, we make use of the linearised induction equation, neglecting the magnetic diffusivity: 
\begin{equation}
    \frac{\partial \boldsymbol{b}}{\partial t} = \nabla \times (\v \times \B).
    \label{eq:induction-lin}
\end{equation}

\noindent Eqs. (\ref{eq:navier-lin}) - (\ref{eq:induction-lin}) can be combined in order to obtain a single equation for $v_z$, the vertical component of $\v$ \citep[see e.g.][for more details regarding this derivation]{chandrasekhar_hydrodynamic_1961}: 

\begin{equation}
    \left(s^2\nabla^2 + g \alpha \beta - \frac{(\B \cdot \nabla)^2}{\mu_0 \rho} \right) \left(s^2 - \frac{(\B \cdot \nabla)^2}{\mu_0 \rho}\right)v_z \quad = -(2 \Om \cdot \nabla)^2 s^2 v_z,
\end{equation}

\noindent where we have assumed that the vertical velocity behaves as $\exp({st})$, $s$ being the growth rate for the convective instability. Furthermore, we assume impenetrable and stress-free boundary temperature conditions, which require that the vertical wavenumber is $k_z = n \pi/\ell_0$, with $n$ an integer. This equation then yields the dispersion relation that relates $s$ to the wavevector $\boldsymbol{k}$:

\begin{equation}
    s^4+s^2\left(2 \omega_B^2+\omega_{\Omega}^2-N^2\right)+\omega_B^4-\omega_B^2 N^2=0,
    \label{eq:dispersion_relation}
\end{equation}

\noindent where : 
\begin{align}
& \omega_B^2 \equiv \frac{(\mathbf{k} \cdot \mathbf{B})^2}{\mu_0 \rho}, \\
& N^2 \equiv \frac{g \alpha \beta k_{\perp}^2}{k^2}, \\
& \omega_{\Omega}^2 \equiv \frac{(2 \Omega \cdot \mathbf{k})^2}{k^2}.
\end{align}

\noindent Following the method in \cite{augustson_model_2019}, we make the previous equation non-dimensional, introducing the following quantities: 
\begin{align}
\label{eq:nondimension_1}
    & N_{*}^2 = \lvert g \alpha \beta \rvert, \\
    & \hat{s} = \frac{s}{N_{*}}, \\
    & z^3 = 1 + a^2 = \frac{k^2}{k_z^2}, \text{ where } a^2 = \frac{k_x^2}{k_z^2} + \frac{k_y^2}{k_z^2} = a_x^2 + a_y^2.
\label{eq:nondimension_2}
\end{align}
In this framework, the dispersion relation given in Eq. (\ref{eq:dispersion_relation}) becomes: 
\begin{equation}
    \hat{s}^4 + \hat{s}^2 \left(2 \mathcal{P}^2 + \mathcal{O}^2 - \frac{(z^3-1)}{z^3} \right) + \mathcal{P}^4 - \mathcal{P}^2 \frac{(z^3-1)}{z^3} = 0,
    \label{eq:dispersion_relation_normalised}
\end{equation}
where the rotation and magnetic parameters are respectively: 
\begin{equation}
    \begin{aligned}
    &\mathcal{O}^2 = \frac{4 \Omega^2}{N_{*}^2},\\ 
    &\mathcal{P}^2 = \frac{\omega_B^2}{N_{*}^2}.
\end{aligned}
\label{eq:o_p_parameters}
\end{equation}

For now on, we suppose that the magnetic field and the rotation axis are aligned with gravity and therefore both vertical, along the $z$ direction, so that : 
\begin{align}
    \vc{\Omega} &= \Omega \vc{e}_z,\\
    \vc{B} &= B \vc{e}_z.
\end{align}
This situation corresponds to a poloidal magnetic field located at the pole of a star or a planet. In this case, the magnetic pulsation is : 
\begin{equation}
    \omega_B^2 = \frac{(k_z B_0)^2}{\mu_0 \rho}.
\end{equation}
This situation is a first step to including both rotation and magnetic fields in MLT, which is suitable for implementation in most stellar evolution codes, which are unidimensional. Considering different geometries that correspond to more complex magnetic field topologies is of interest for 2D stellar evolution codes such as ESTER \citep{rieutord_algorithm_2016}. Such configurations are left for future work. 

\subsection{Heat-flux maximisation}

We use the method defined by \cite{malkus_heat_1954}, which relies on the heat-flux maximisation principle. We assume that the linear mode maximising the total convective heat flux is dominant. When linking the theory with the Mixing-Length Theory, this mode is selected to compute the characteristic velocity and wavenumber for the flow. In the absence of any diffusive process, the convective heat-flux writes \citep[see e.g.][]{stevenson_turbulent_1979}: 
\begin{equation}
    \mathcal{F}_{\Omega,B} = \frac{\rho c_p}{\alpha g} \frac{s^2}{k^2} \left(s + \frac{\omega_B^2}{s} \right).
    \label{eq:flux}
\end{equation}
The first term in Eq. (\ref{eq:flux}) corresponds to the kinetic energy, and the second to the magnetic energy. 
In this whole article, the subscript $\cdot_{\Omega,B}$ denotes the rotating and magnetised case, while the subscript $\cdot_0$ refers to the non-rotating and non-magnetised situation in standard MLT. 
Using the notations defined in Eq. (\ref{eq:nondimension_1})-(\ref{eq:nondimension_2}), it boils down to : 

\begin{equation}
    \mathcal{F}_{\Omega,B} = \frac{\rho c_p}{\alpha g} \frac{N_{*}^3}{k^2} \hat{s} \left(\hat{s} + \frac{\mathcal{P}^2}{\hat{s}} \right).
    \label{eq:heat_flux}
\end{equation}

We compare with this method the convective values with rotation and magnetic fields to the ones without rotation or magnetic field. In the standard non-rotating non-magnetised case, the dominant mode verifies: 
\begin{equation}
    k_0^2 = \frac{5}{2}k_z^2,
\end{equation}
where $k_0$ is the convective wavenumber in standard MLT.
In this framework, the growth rate of the most unstable mode is:
\begin{equation}
    s_0^2 = \frac{3}{5}g_0 \alpha \beta_0, 
    \label{eq:s_0}
\end{equation}
where $\boldsymbol{g}_0$ (resp. $\beta_0$) is the effective gravity (resp. thermal gradient) in the non-rotating and non-magnetised case.
Finally, the convective flux without rotation or magnetism is: 
\begin{equation}
    \mathcal{F}_0 = \mathcal{N}_0\frac{\rho c_p}{\alpha g k_z^2} N_{*, 0}^3,
\label{eq:flux_0}
\end{equation}
where $\mathcal{N}_0$ is a dimensionless prefactor: 
\begin{equation}
    \mathcal{N}_0 = \left(\frac{3}{5}\right)^{3/2} \frac{2}{5}.
\end{equation}

We then consider the variation of superabadiaticity for this system:
\begin{equation}
    \epsilon = \frac{H_p \beta}{T},
\end{equation}
where $H_p$ is the pressure scale height. The superadiabaticity is the difference between the temperature gradient and the adiabatic temperature gradient. The more efficient the convection, the closer the temperature gradient is to the adiabatic temperature gradient, and the lower the superadiabaticity. Regions with less efficient convection then have higher values of superadiabaticity \citep[e.g.][]{sabhahit_superadiabaticity_2021}
In turn, we have: 
\begin{equation}
    N_{*}^2 = \left \lvert g \alpha T \frac{\epsilon}{H_p} \right \rvert.
\end{equation}
We must then introduce the ratio of superabadiaticities as an additional unknown: 
\begin{equation}
    q = \frac{N_{*,0}}{N_{*}}.
    \label{eq:def_q}
\end{equation}
This leads to the modulation of superadiabaticity in the rotating and/or magnetised case, compared to the non-rotating and non-magnetised situation:
\begin{equation}
    \tilde{\epsilon} = \frac{\epsilon_{\Omega,B}}{\epsilon_0} = \frac{1}{q^2}.
\label{eq:ratio_eps}
\end{equation}
Within this framework, we compute the ratio of the convective wavenumber and velocity in the presence of rotation and/or a magnetic field to their counterparts in the non-rotating and non-magnetised case.

\begin{equation}
    \tilde{K} = \frac{k_{\Omega,B}}{k_0} = z^{3/2} \sqrt{\frac{2}{5}},
    \label{eq:ratio_k}
\end{equation}
\begin{equation}
    \tilde{U} = \frac{v_{\Omega,B}}{v_0} = \frac{s k_0}{k s_0}=z^{-3/2} \frac{5 \hat{s}}{\sqrt{6} q}.
\label{eq:ratio_v}
\end{equation}

\subsection{Rossby and Alfvén number}

As in the previous works by \cite{stevenson_turbulent_1979} and \cite{augustson_model_2019}, we give an expression for the modification of the convective velocity and the convective wavenumber compared to the non-rotating and non-magnetised case depending on the dimensionless inverse Alfvén number and Rossby number. 
Following Eq. (\ref{eq:s_0}), the convective velocity without rotation or magnetism is : 
\begin{equation}
    v_0 = \frac{s_0}{k_0} = \frac{\sqrt{6}}{5} \frac{N_{*,0}}{k_z}.
    \label{eq:v0_def}
\end{equation}
Moreover, the convective vertical wavenumber is constrained by the boundary conditions: 

\begin{equation}
    k_z = \frac{\pi}{\ell_0},
\end{equation}
$\ell_0$ being the vertical size of the studied system. 

In this framework, the Rossby number is the ratio between the advection and the Coriolis acceleration in the Navier-Stokes equation.
\begin{equation}
    \mathcal{R}o = \frac{v_0}{2 \ell_0 \Omega},
\end{equation}
That yields with Eq. (\ref{eq:v0_def}): 
\begin{equation}
    \mathcal{R}o  = \frac{\sqrt{6} N_{*,0}}{10 \pi \Omega}.
\end{equation} 
Finally, as in \cite{augustson_model_2019}, we derive the relation between the rotation parameter $\mathcal{O}$ from Eq. (\ref{eq:o_p_parameters}) and $\mathcal{R}o$: 

\begin{equation}
    \mathcal{R}o = \frac{\sqrt{6}q}{5 \pi \mathcal{O}}.
    \label{eq:rossby}
\end{equation}

The same can be done to determine the relation between the inverse Alfvén number and the magnetic parameter $\mathcal{P}$ from Eq.(\ref{eq:o_p_parameters}). The inverse Alfvén number is the ratio between the Alfvén velocity $v_A$ and the convective velocity $v_0$: 
\begin{equation}
    A = \frac{v_A}{v_0}, 
\end{equation}
where $v_A = \sqrt{\frac{B^2}{\mu_0 \rho}}$. 
In the case of a magnetic field along the $z-$axis, one has : 
\begin{equation}
    \mathcal{P}^2 = \frac{k_z^2 B^2}{\mu_0 \rho N_{*,0}^2}.
\end{equation}
Finally, we derive the expression for the Alfvén number: 
\begin{equation}
    A = \frac{5}{\sqrt{6}} \frac{\mathcal{P}}{q}.
    \label{eq:alfven}
\end{equation}

\subsection{Invariance of the heat-flux}
Following the previous sections, we are left with two equations (\ref{eq:heat_flux}) and (\ref{eq:dispersion_relation_normalised}) to determine the wavenumber that maximises the heat flux. With these equations, one can determine $\hat{s}$ and $z$.

However, determining the convective velocity and the convective wavenumber requires knowledge of the superabadiaticities ratio $q$. To do so, we need an additional hypothesis on the system. As in the previous work by \cite{augustson_model_2019}, we assume that $\text{max} [\mathcal{F}_0] = \text{max} [\mathcal{F}]$, i.e. the heat-flux with rotation and magnetism is equal to the maximum value of the heat flux without rotation nor magnetism. This is based on the assumption that the total heat-flux should not change with rotation or magnetic field, as it is determined from the nuclear reaction energy rate inside stars or the radioactive and chemical energy in planets. However, one has to note that it is not true in general for any convective system.
\noindent In this framework, we have: 
\begin{equation}
    \text{max}[\mathcal{F}_0] = \text{max}[\mathcal{F}_{\Omega,B}].
\end{equation}
Using Eqs. (\ref{eq:heat_flux})-(\ref{eq:flux_0}), it yields: 
\begin{equation}
    \frac{\rho c_p}{\alpha g} \frac{N_{*}^2}{k^2} \hat{s}^2 \left( \hat{s} + \frac{\mathcal{P}^2}{\hat{s}} \right) = \frac{\rho c_p}{\alpha g} \frac{N_{*}^2}{k_z^2} \hat{s}_0^3.
\end{equation}
Using Eq. (\ref{eq:def_q}) and (\ref{eq:def_q})), we find: 
\begin{equation}
    \mathcal{N}_0 q^3 z^3 = \hat{s}^2 \left(\hat{s} + \frac{\mathcal{P}^2}{\hat{s}}\right).
    \label{eq:heat_flux_equal}
\end{equation}
Finally, this theoretical model consists of four steps: 
\begin{enumerate}
    \item Defining the dispersion relationship that links $\hat{s}$ and $z$ (Eq. \ref{eq:dispersion_relation_normalised}).
    \item Maximising the heat-flux given by Eq. (\ref{eq:heat_flux}) for given parameters $\mathcal{O}$ and $\mathcal{P}$. One then finds a couple of values $(\hat{s}_m, z_m)$ that make the heat flux maximal. 
    \item Determining the value of $q_m$ with the help of Eq. (\ref{eq:heat_flux_equal}). 
    \item Computing the corresponding Rossby number and/or Alfvén number, using Eq.(\ref{eq:alfven}) and Eq. (\ref{eq:rossby}).
    \item One can in turn determine the ratio of convective velocities $\tilde{U}$, convective wavenumber $\tilde{K}$ and superadiabaticities $\tilde{\epsilon}$ compared to the non-rotating and non-magnetised case. For this, we use Eqs. (\ref{eq:ratio_v}), (\ref{eq:ratio_k}) and (\ref{eq:ratio_eps}).
\end{enumerate}
Moreover, one can note that Eq. (\ref{eq:heat_flux_equal}) combined with Eqs. (\ref{eq:ratio_eps})-(\ref{eq:ratio_v}) yields the conservation of the following quantity on $\tilde{K}$ and $\tilde{U}$:
\begin{equation}
    \frac{\tilde{U}}{\tilde{K}} \left( \tilde{U}^2 \tilde{K}^2 + \frac{2 A^2}{5} \right) = 1,
    \label{eq:conservation_flux}
\end{equation}
where the first term is a kinetic energy flux and the second term is linked to the magnetic energy. 
In the following section, we apply this methodology to three distinct cases. We first analyze the impact of rotation and magnetic fields on convection separately before examining their combined effect.

\section{Rotating convection}
\label{sec:rmlt}

\subsection{Prior studies of rotating MLT}
The rotating MLT model has been developed by \cite{stevenson_turbulent_1979}, based on the heat-flux maximisation principle from \cite{malkus_heat_1954}. The author provides some prescriptions for the modification of the convective velocity (resp. wavenumber and superadiabaticity) for the rapid-rotation and the slow-rotation limits, which depend on the local Rossby number $\mathcal{R}o$. Table \ref{tab:stevenson_rot} shows the expressions for $\tilde{U}_\Omega \equiv v_\Omega/v_0$, $\tilde{K}_\Omega \equiv k_\Omega/k_0$ and $\tilde{\epsilon}_{\Omega} \equiv \epsilon_\Omega/\epsilon_0$ from \cite{stevenson_turbulent_1979}, for the high-Rossby number and low-Rossby number cases, respectively. The lower the Rossby number, the lower the convective velocity, and the higher the convection wavenumber. This is in agreement with the general tendency that rotation tends to stabilise convection, making it less vigorous, as discussed in Sec. \ref{sec:rotmag} \citep[e.g.][]{chandrasekhar_hydrodynamic_1961}. 

More recently, \cite{augustson_model_2019} extended this model to tackle the intermediate Rossby numbers ranges, including the viscosity and heat diffusion. Their approach is based on solving the following equation, which gives access to $z_m$ such that the convective heat-flux is maximised: 
\begin{equation}
    2 z_m^5 - 5 z_m^2 - \frac{18}{25 \pi^2 \mathcal{R}o^2 \mathcal{N}_0^{3/2}} = 0.
    \label{eq:equation_kyle}
\end{equation}
The resulting convective velocity (resp. wavenumber) is then computed with Eq. (\ref{eq:ratio_v}) (resp. Eq. \ref{eq:ratio_k}). The tendency for the high-rotation case is the same as in \cite{stevenson_turbulent_1979}.

\begin{table}[h!]
    \centering
    \caption{Modulation of the convective velocity, wavenumber and superadiabaticity by rotation in the rapidly- and slowly-rotating regimes, respectively, following \cite{stevenson_turbulent_1979}.}
\label{tab:stevenson_rot}
    \renewcommand{\arraystretch}{2}
    \begin{tabular}{p{1.5cm}  p{2.5cm}  p{2.5cm}}
    \hline
     \hline
     & $\mathcal{R}o \ll 1$ &  $\mathcal{R}o \gg 1$ \\
     \hline
        $\tilde{U}_\Omega (\mathcal{R}o)$ & $1.5 \mathcal{R}o^{1/5}$   & $1 - \frac{1}{242 \mathcal{R}o^2}$\\
        
         $\tilde{K}_\Omega(\mathcal{R}o)$ & $0.5 \mathcal{R}o^{-3/5}$   & $1+ \frac{1}{82 \mathcal{R}o^2}$\\

          $\tilde{\epsilon}_\Omega (\mathcal{R}o)$ & $0.23 \mathcal{R}o^{-4/5}$   & $1+ \frac{1}{62 \mathcal{R}o^2}$\\
    \hline  
    \end{tabular}
\end{table}

Some numerical simulations of rotating convection have given interesting results compared with R-MLT \citep{kapyla_local_2005, barker_theory_2014, currie_convection_2020}. \cite{barker_theory_2014} tested this theory with three-dimensional non-linear, hydrodynamical simulations of Boussinesq convection in a Cartesian box. The rapid rotation regime established in \cite{stevenson_turbulent_1979} seems to hold well for almost three decades in Rossby number (from $\mathcal{R}o \sim 10^{-1}$ to $\mathcal{R}o \sim 10^{-4}$ approximately).  \cite{currie_convection_2020} then extended this work to tackle the case of a rotation axis misaligned with gravity. \cite{stevenson_turbulent_1979} prescriptions of Rotating Mixing-Length Theory (R-MLT) agree well with the simulations, although it fails to capture the anisotropy of rotating convection. \cite{augustson_model_2019} method also compares well to those simulations, although prefactor for the wavenumber for the low-Rossby differs from the result in \cite{stevenson_turbulent_1979} (see Fig. \ref{fig:rmlt}). 

The R-MLT framework has been of great use in explaining a wide variety of phenomena taking place in stars' and planets' convective zones. \cite{augustson_model_2019} also derived a prescription for the convective overshoot in the presence of rotation, which has been compared successfully to 3D non-linear spherical numerical simulations \citep{korre_dynamics_2021}. The R-MLT also provides insightful results to understand light-elements mixing in late-type stars \citep{dumont_lithium_2021}, as well as the properties of the convective core boundary in early-type stars \citep{michielsen_probing_2019}. Furthermore, R-MLT prescriptions from \cite{stevenson_turbulent_1979} have been implemented to model the stellar structure of fully convective stars in \cite{ireland_radius_2018} with rotation. It was shown that the entropy gradient of such stars is modified by rapid rotation in the bulk of their interior, thus leading to radius inflation and evolutionary changes. However, this impact on stellar structure is found to be negligible, as the overall structure is primarily influenced by the outer layers, where rotation has little effect on convection. In addition, this model was then used to quantify the excitation of waves by turbulent convection in the presence of rotation for gravito-inertial modes \citep{mathis_impact_2014,augustson_model_2020}, and acoustic modes in solar-like stars \citep{bessila_impact_2024}. In this framework, it was used to explain the mixing by waves in equatorial numerical simulations of rotating intermediate-mass stars \citep{varghese_effect_2024}. When it comes to gaseous giant planets, \cite{fuentes_rotation_2023} showed that rotation significantly decreases the mixing of heavy elements efficiency in such planets, which could explain why Jupiter and Saturn are not fully mixed. Finally, the R-MLT framework is also of interest to assess tidal dissipation in stars or giant planets, taking into account rotation \citep{mathis_impact_2016,  ahuir_dynamical_2021, devries_tidal_2023}. It was shown that rotation decreases the eddy-viscosity in the rapid rotation regime, which leads to a decrease in efficiency for the dissipation of the equilibrium tide, while it is amplified for the dynamical tide.\\

\subsection{Rotating Mixing-Length Theory}

\begin{figure}[h]
    \centering
    \includegraphics[width=\linewidth]{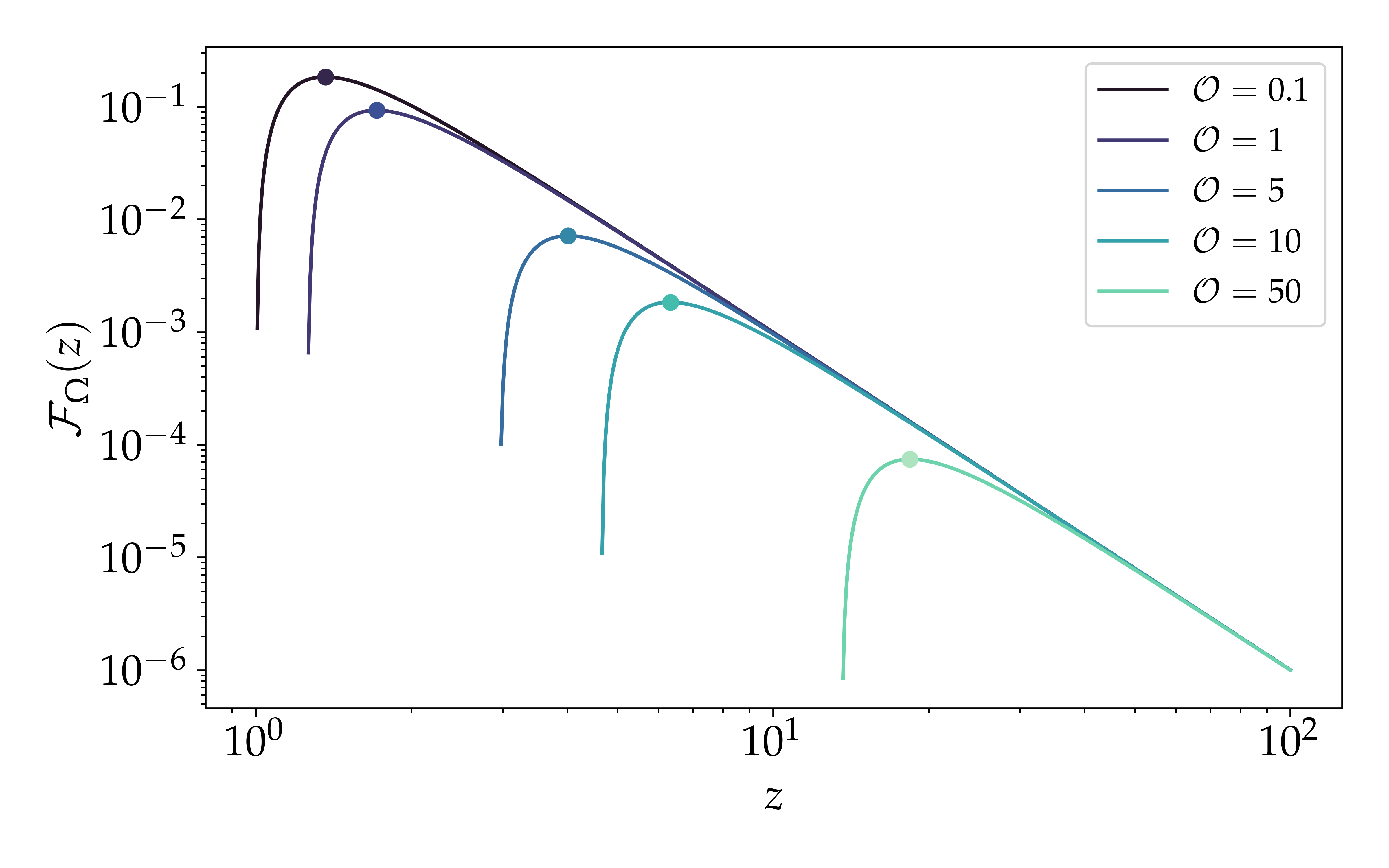}
    \caption{Convective heat-flux from Eqs. (\ref{eq:dispersion_rot}) - (\ref{eq:heat_flux_rot}) for different values of the rotation parameter $\mathcal{O}$. The dots indicate the maxima for each value of $\mathcal{O}$.}
    \label{fig:heat_flux_rot}
\end{figure}

In this first section, we consider non-magnetised convection with $\Omega \neq 0$ and compare our results to the ones from \cite{stevenson_turbulent_1979} and \cite{augustson_model_2019}. To do so, we take the limit when $\mathcal{P}=0$ in equations (\ref{eq:dispersion_relation_normalised}), (\ref{eq:heat_flux}) and (\ref{eq:heat_flux_equal}).
\begin{enumerate}
    \item The dispersion relation becomes: 
    \begin{equation}
\hat{s}^2 + \mathcal{O}^2 - \frac{(z^3-1)}{z^3}   = 0.
\label{eq:dispersion_rot}
\end{equation}

\item The convective heat-flux, that we maximise numerically is: 
\begin{equation}
    \mathcal{F}_{\Omega} = \frac{\rho c_p}{\alpha g} \frac{N_{*}^3}{k^2} \hat{s}^3.
    \label{eq:heat_flux_rot}
\end{equation}
\item The hypothesis of an invariant heat-flux gives:
\begin{equation}
    \mathcal{N}_0 q^3 z^3 = \hat{s}^3.
\end{equation}
\end{enumerate}
This set of equations allows us to determine the values of $z_m, \hat{s}_m, q_m$ such that the heat-flux is maximum. We then make use of Eq.(\ref{eq:rossby}) to compute the corresponding Rossby number and Eqs. (\ref{eq:ratio_k})-(\ref{eq:ratio_v}) to calculate the modulation of the convective velocity and convective wavenumber. The heat-flux computed with Eq. (\ref{eq:heat_flux_rot}) is plotted on Fig. \ref{fig:heat_flux_rot} for different values of $\mathcal{O}$. When the rotation rate is higher, $\mathcal{O}$ is higher, and the overall heat-flux diminishes. The reduced wavenumber $z_m$ for which the flux is maximal is then higher. Directly from Eq.(\ref{eq:ratio_k}), the convective wavenumber is higher, which means the convective length scale is lower.

\begin{figure}[h]
    \centering
    \includegraphics[width=\linewidth]{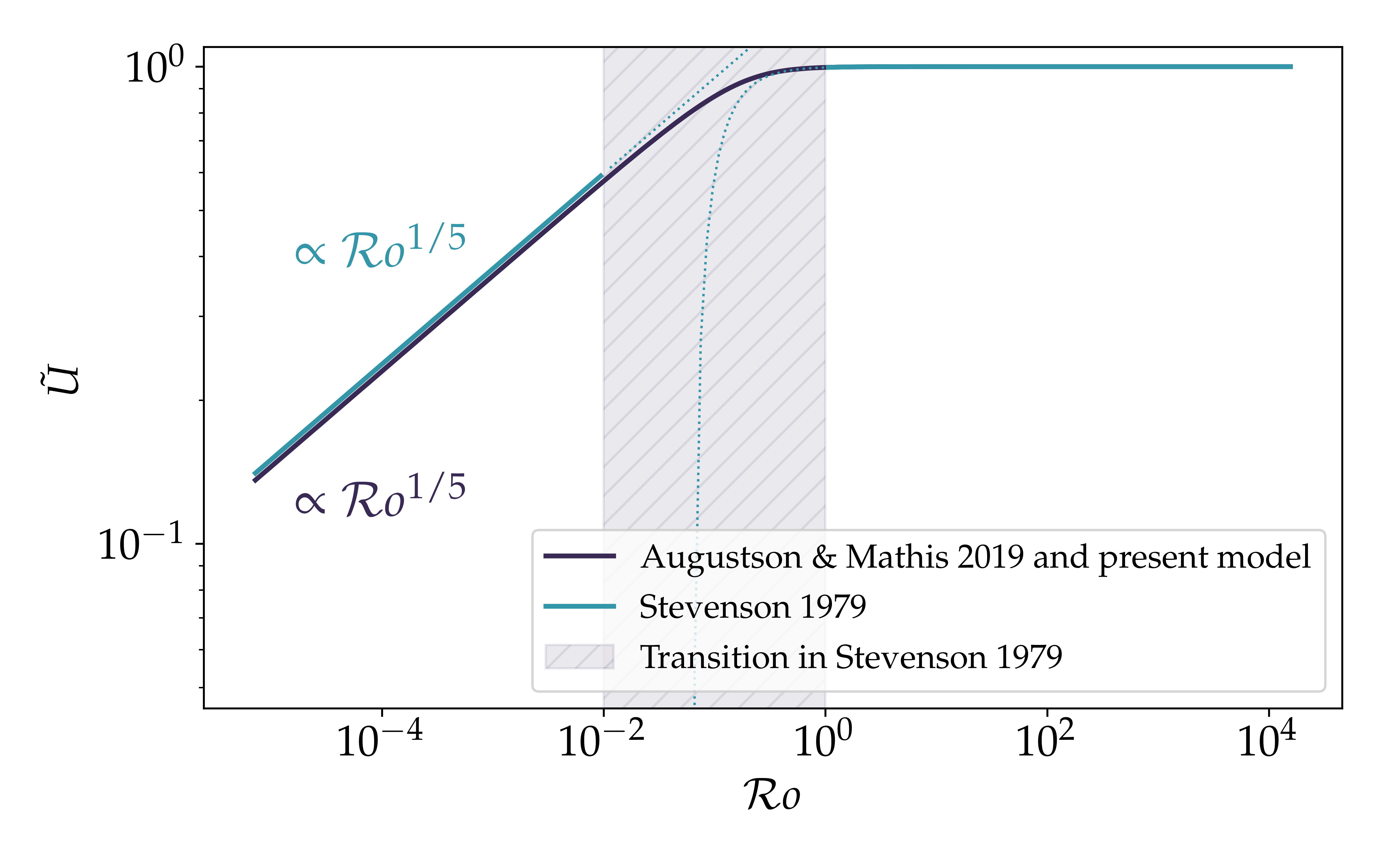}
        \includegraphics[width=\linewidth]{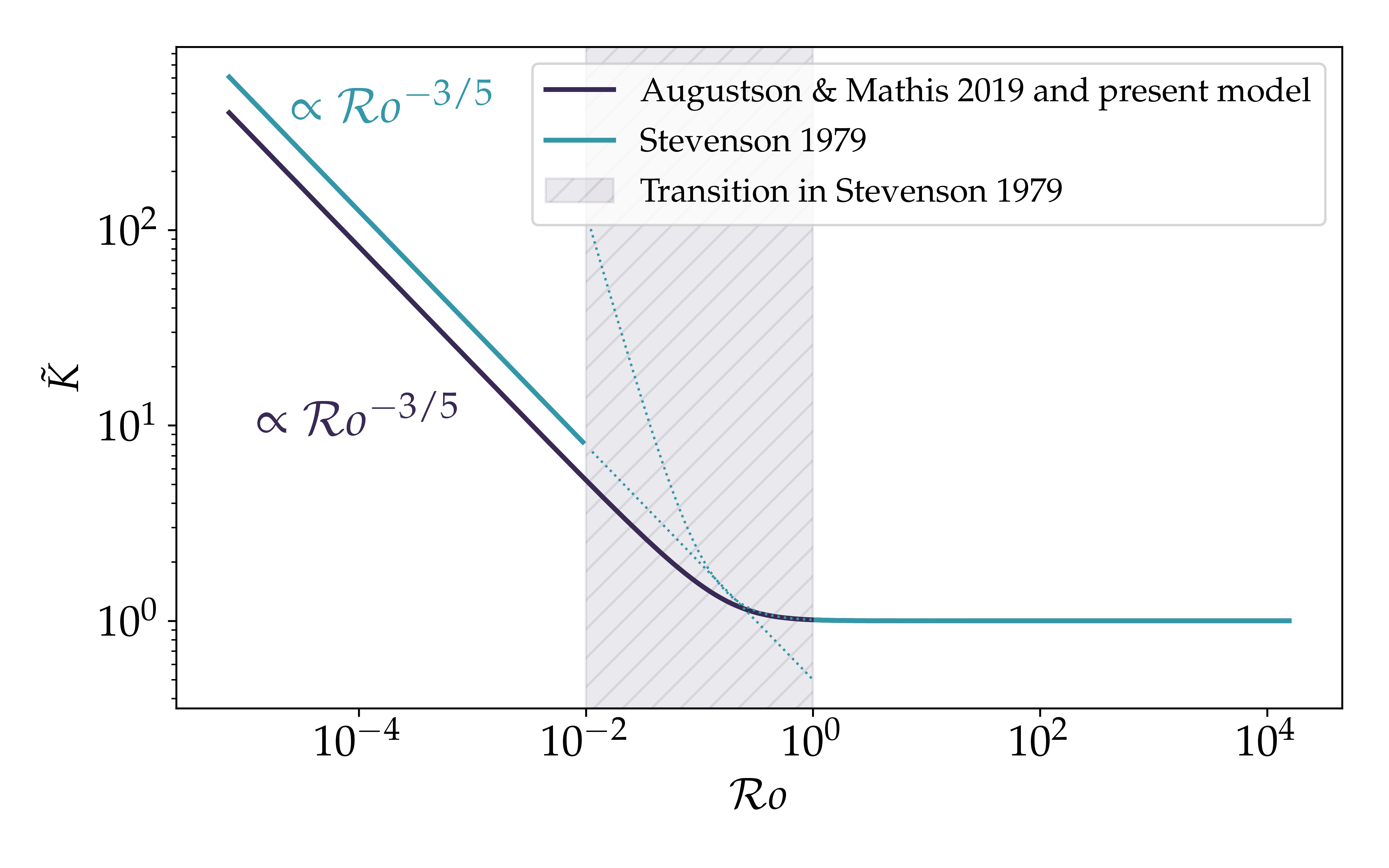}
    \includegraphics[width=\linewidth]{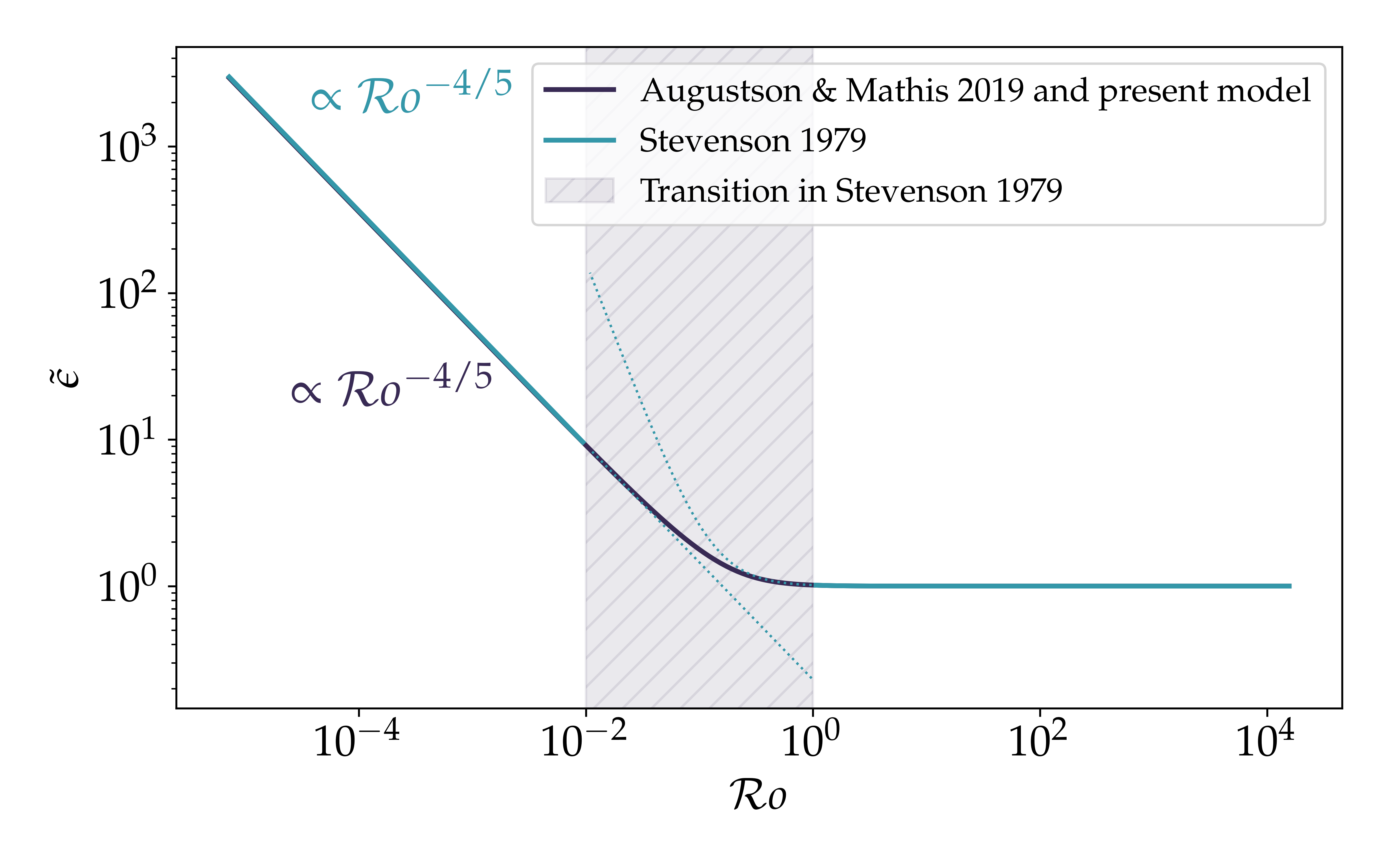}
    \caption{R-MLT results from \cite{stevenson_turbulent_1979}, \cite{augustson_model_2019} and the present work. The results from our work and \cite{augustson_model_2019} are in close agreement. Top: Convective velocity modulation with rotation $\tilde{U}$ as a function of the Rossby number. Center: Convective wavenumber modulation with rotation $\tilde{K}$ as a function of the Rossby number. Bottom: Superabadiaticity modulation with rotation $\tilde{\epsilon}$ as a function of the Rossby number.}
    \label{fig:rmlt}
\end{figure}

The results for the modified convective velocity $\tilde{U}$, the modified wavenumber $\tilde{K}$ and the modified superadiabaticity $\tilde{\epsilon}$ are plotted in Fig. \ref{fig:rmlt} and compared with previous works from \cite{stevenson_turbulent_1979}, \cite{augustson_model_2019}. In all cases, the same trend emerges: as the rotation rate increases, the convective velocity decreases, while the convective wavenumber and superadiabaticity increase. This aligns with the general tendency that higher rotation leads to less efficient convection, resulting in smaller convective length scales and lower velocities. The increase in superadiabaticity further indicates reduced convective efficiency.
The results from our work and the ones from \cite{augustson_model_2019} are the same, with the corresponding curves perfectly overlapping in Fig.\ref{fig:rmlt}. Indeed, the governing equations are identical, while there is a slight difference in the computation method. \cite{augustson_model_2019} provides a fully analytical solution with their Eq. (46) (which is Eq. \ref{eq:equation_kyle} of the present paper), computing the maximum of the heat flux by differentiating Eqs. (\ref{eq:heat_flux_rot})-(\ref{eq:dispersion_rot}) with respect to $z$. In the present study, we compute the maximum of the heat-flux numerically, as shown in Fig. \ref{fig:heat_flux_rot}. 
When compared to the asymptotic results from \cite{stevenson_turbulent_1979}, the convective velocity modulation computed in \cite{augustson_model_2019} and our work agree for the rapid-rotation case and the slow-rotation case. However, there is a difference in prefactor for the convective wavenumber modulation for the low Rossby number case. Both authors considered the same linear dispersion relation for the convective instability, but there can subsist a difference in the prefactor. For the rapid-rotation asymptotic regime when $\mathcal{R}o \lesssim 10^{-2}$, we find the following scalings:

\begin{equation}
    \tilde{U}_\Omega (\mathcal{R}o) \sim 1.422 \cdot \mathcal{R}o^{0.197},
\end{equation}

\begin{equation}
    \tilde{K}_\Omega (\mathcal{R}o) \sim 0.329 \cdot \mathcal{R}o^{-0.6000},
\end{equation}

\begin{equation}
    \tilde{\epsilon}_\Omega (\mathcal{R}o) \sim 0.226 \cdot \mathcal{R}o^{-0.8000}.
\end{equation}

These scalings show good agreement with the prescriptions of \cite{stevenson_turbulent_1979} from Tab. \ref{tab:stevenson_rot}. Both the exponents and prefactors generally align, with one exception: the prefactor for the wavenumber modulation, which is 0.329 in our study (similar to \cite{augustson_model_2019}) compared to 0.5 in \cite{stevenson_turbulent_1979}. The origin of this discrepancy remains unexplained. In the following section, we focus on the impact of magnetism on convection in the framework of MLT.

\section{Magnetised convection}
\label{sec:mmlt}
\subsection{Prior studies of magnetised convection}
\label{sec:studies_magnetic}
In stellar physics modelling, the most used approach to tackle magnetised convection is to study the stability criterion. Without rotation nor magnetism, \cite{schwarzschild_equilibrium_1906} and \cite{ledoux_stellar_1947} derived instability criteria for convection to start, which are used in stellar structure and evolution codes. As explained in the introduction, convection is stabilised by magnetic fields through the Lorentz force: if the imposed magnetic field is too strong, the fluid becomes stable to convection. \cite{gough_influence_1966} extended the previous stability criteria to include the action of an imposed magnetic field, based on an energy principle \citep{bernstein_ib_energy_1958}. Such a stability criterion has been implemented in stellar modelling codes to study M-dwarf stars and brown dwarfs. Indeed, the observed radii of such celestial bodies are consistently larger than standard non-magnetised models' predictions \citep[e.g.][]{mullan_are_2001}. A strong magnetic field would mitigate convection in stellar interiors, then leading to radius inflation \citep[see e.g.][]{mullan_are_2001, mullan_structure_2003, mullan_magnetic_2010, ireland_radius_2018, macdonald_magnetic_2019, mullan_magneto-convective_2023, macdonald_possibility_2024}.
Another application of \cite{gough_influence_1966} criterion in stellar physics is to assess the critical magnetic field $B_{\rm crit}$, which is the minimum magnetic field that freezes convection. It has been useful in explaining magnetic field bimodal distribution in early-type stars \citep{jermyn_origin_2020, farrell_initial_2022}. Strong fossil magnetic fields can effectively freeze convection in subsurface convective zones, allowing the detection of intense fields at the stellar surface \citep{wade_mimes_2016, shultz_magnetic_2019}. In contrast, weaker fossil fields do not suppress convection and remain beneath the surface, unlike dynamo-generated fields that emerge from convection zones.

Note that a complementary tool to study the onset of convection in the presence of a magnetic field is to use the critical Rayleigh number in a linear framework for the convective instability. Convection takes place in systems where the Rayleigh number exceeds the critical value. When the magnetic field is strong, the critical Rayleigh number becomes higher \citep[e.g.][]{chandrasekhar_hydrodynamic_1961, eltayeb_hydromagnetic_1977, horn_elbert_2022}, leading to the same conclusions that \cite{gough_influence_1966}: convection is more difficult to start in magnetised fluids. To our knowledge, stability criteria involving the Rayleigh number in a magnetic framework have not been implemented in the context of stellar structure and evolution modelling, although some studies showed that some HeI subsurface convection zones predicted in stellar evolution codes should have a subcritical Rayleigh number, and therefore no convective motion is expected \citep{jermyn_atlas_2022}.

\begin{figure}[h]
    \centering
        \includegraphics[width=0.8\linewidth]{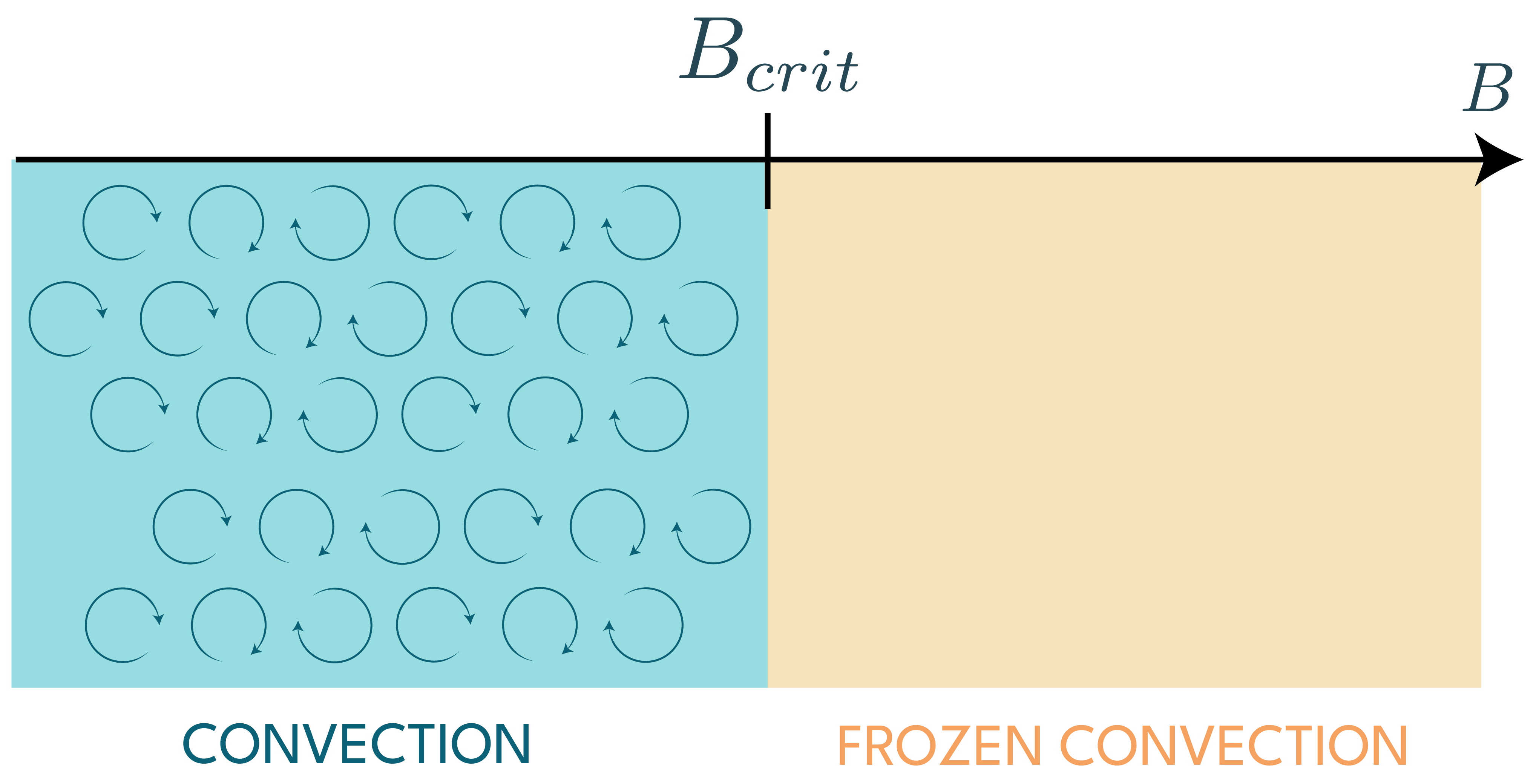}
        \includegraphics[width=0.8\linewidth]{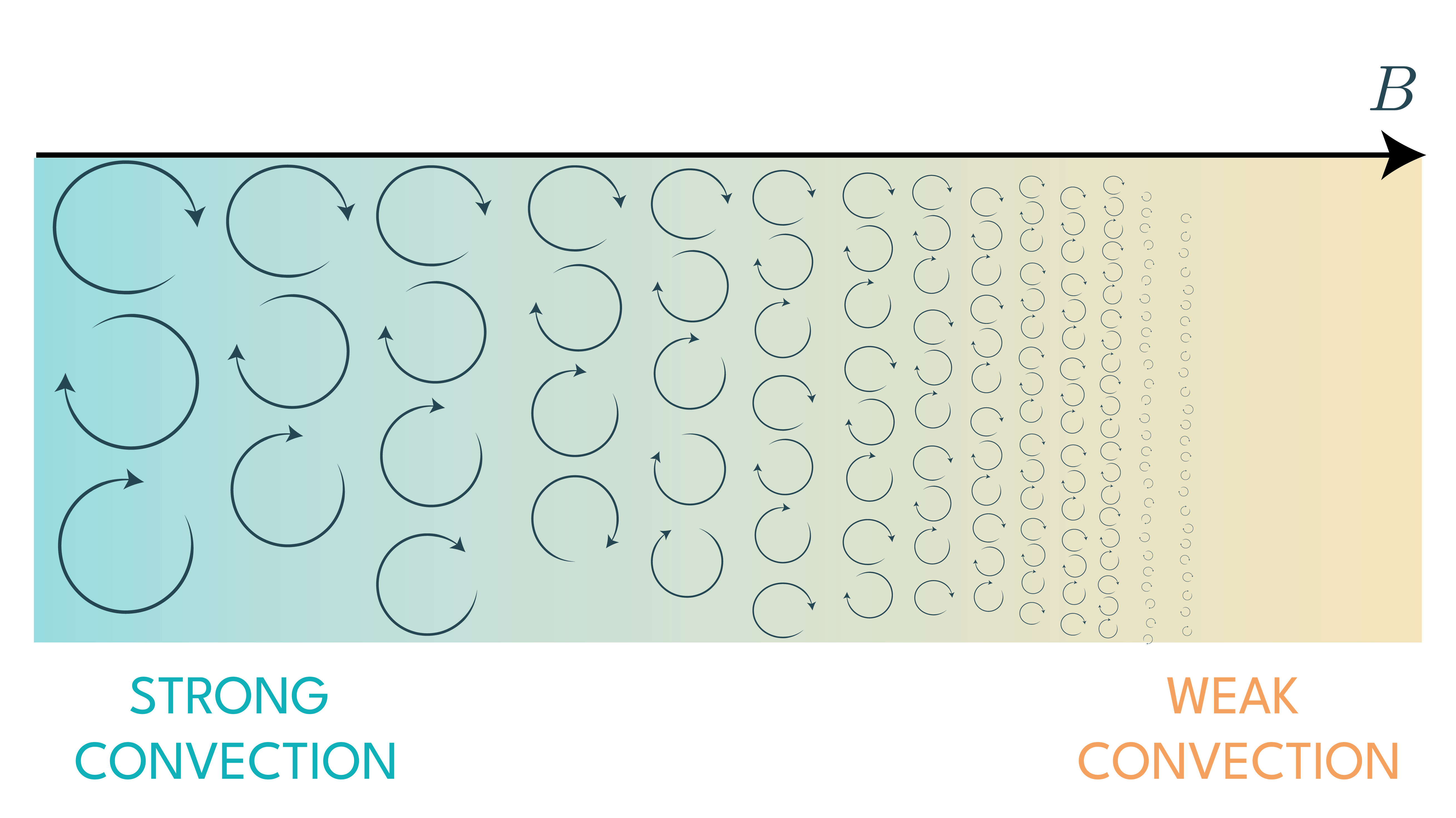}

    \caption{Top: The critical magnetic field approach for magnetised convection. Bottom: The Mixing-Length Theory approach.}
    \label{fig:critical_field}
\end{figure}

Whether it comes from a linear approach \citep{chandrasekhar_hydrodynamic_1961} or an energy principle \citep{gough_influence_1966}, the stability criteria of magnetised convection are "on-off" consideration: either convection is active, or there is no convection. It does not account for the progressive weakening of convection. Following his work on rotating convection, \cite{stevenson_turbulent_1979} also proposed some scaling relations for the modification of the characteristic convective velocity and wavenumber in Magnetised Mixing-Length Theory (M-MLT). Those scalings, detailed in Tab. \ref{tab:stevenson_mag}, depend on the Alfvén number $A$. As opposed to the stability criteria, this shows that the convective velocity diminishes progressively when the magnetic field increases, as well as the convective length scale. (see the illustration Fig. \ref{fig:critical_field}). Unlike the work on rotating convection from the same paper \citep{stevenson_turbulent_1979}, no direct numerical simulations nor observational work has been realised to assess the validity of such prescriptions, although \cite{hotta_breaking_2018} showed that the general tendency agrees qualitatively. It has nevertheless been used to evaluate the impact of magnetic field on the amplitudes of acoustic modes generated by turbulent convection in solar-like stars \citep{bessila_stochastic_2024}. 

\begin{table}[h!]
    \centering
    \renewcommand{\arraystretch}{2.5}
    \begin{tabular}{p{1.5cm}  p{2.5cm}  p{2.5cm}}
    \hline
     \hline
     & $A \ll 1$ &  $A \gg 1$ \\
     \hline
        $\tilde{U}_B(A)$ & $1-\frac{11A}{75}$   & $\frac{0.92}{\sqrt{A}}$\\
        
         $\tilde{K}_B(A)$ & $1 + \frac{A}{25}$   & $0.49 \sqrt{A}$\\

          $\tilde{\epsilon}_B(A)$ & $1 + \frac{2A}{15}$   & $0.24 A$\\
    \hline  
    \end{tabular}
    \caption{Values of the modulation factors $\tilde{U}(A)$ and $\tilde{K}(A)$ in the magnetised case \citep{stevenson_turbulent_1979}.}
    \label{tab:stevenson_mag}
\end{table}

\subsection{Magnetised Mixing-Length Theory (M-MLT)}

Here, we consider non-rotating magnetised convection with $\Omega = 0$ and compare our results to the ones derived by \cite{stevenson_turbulent_1979}. To do so, we take the limit when $\mathcal{O}=0$ in the set of equations (\ref{eq:dispersion_relation_normalised}), (\ref{eq:heat_flux}) and (\ref{eq:heat_flux_equal}).

\begin{enumerate}
    \item The dispersion relation becomes: 
    \begin{equation}
\hat{s}^4 + \hat{s}^2 \left(2 \mathcal{P}^2 - \frac{(z^3-1)}{z^3} \right) + \mathcal{P}^4 - \mathcal{P}^2 \frac{(z^3-1)}{z^3} = 0.
\label{eq:dispersion_mag}
\end{equation}

\item The convective heat-flux, that we maximise numerically is: 
\begin{equation}
     \mathcal{F}_{B} = \frac{\rho c_p}{\alpha g} \frac{\hat{s}^2}{k^2} \left(s + \frac{\omega_B^2}{\hat{s}} \right).
    \label{eq:heat_flux_mag}
\end{equation}
\item The hypothesis of an invariant heat-flux gives:
\begin{equation}
    \mathcal{N}_0 q^3 z^3 = \hat{s}^2 \left(\hat{s} + \frac{\mathcal{P}^2}{\hat{s}}\right).
\end{equation}
\end{enumerate}

\begin{figure}[h!]
    \centering
    \includegraphics[width=\linewidth]{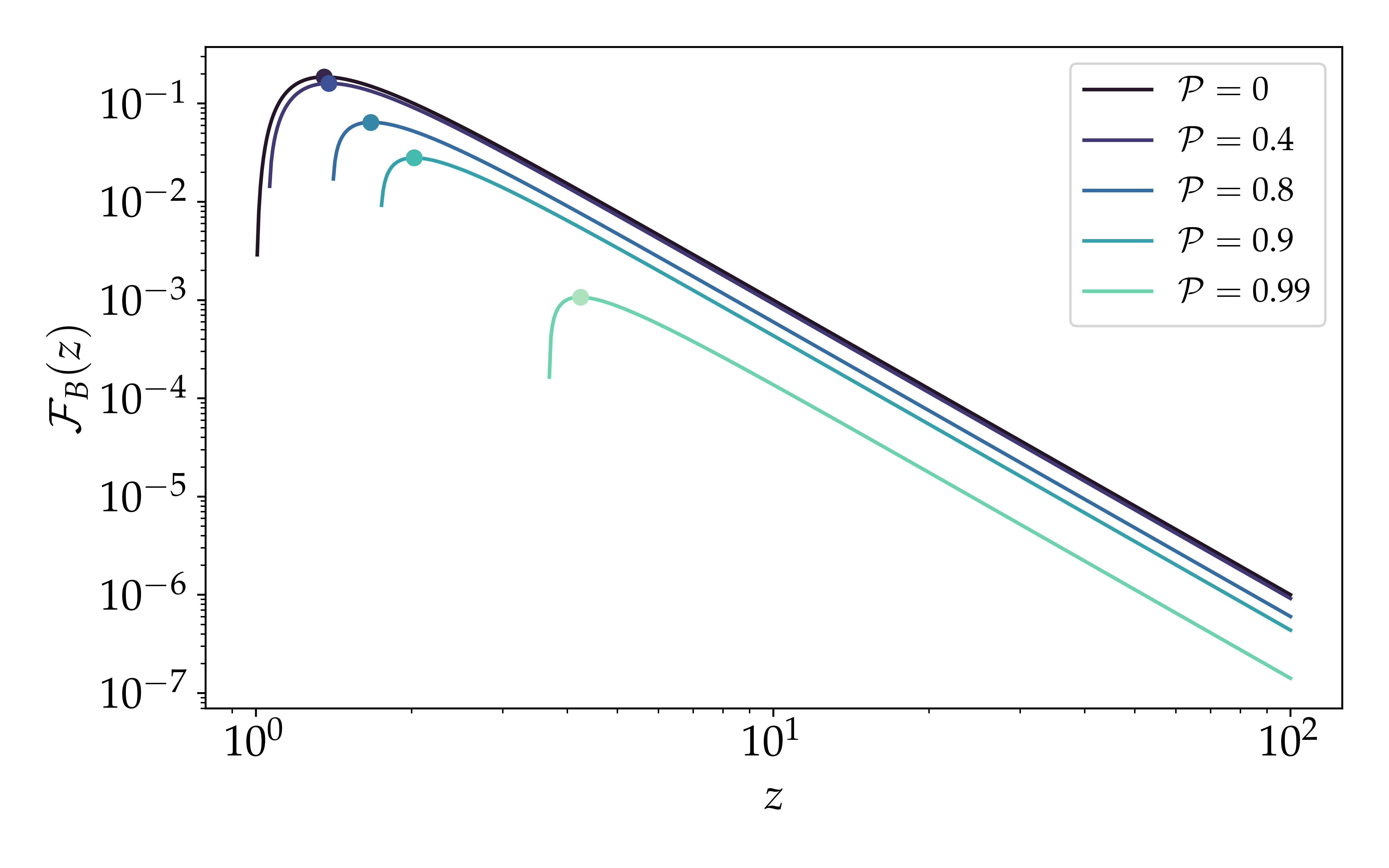}
    \caption{Convective heat-flux from Eqs. (\ref{eq:dispersion_mag}) - (\ref{eq:heat_flux_mag}) for different values of the magnetic parameter $\mathcal{P}$. The dots indicate the maxima for each value of $\mathcal{P}$.}
    \label{fig:heat_flux_mag}
\end{figure}

As for the rotating case in Sec. \ref{sec:rmlt}, this set of equations allows us to determine $z_m, \hat{s}_m, q_m$ which makes the heat-flux maximum. We then make use of Eq.(\ref{eq:alfven}) to compute the corresponding Alfvén number and Eqs. (\ref{eq:ratio_k})-(\ref{eq:ratio_eps}) to calculate the modulation of the convective velocity, wavenumber and superabadiaticity. The heat-flux computed with Eqs. (\ref{eq:dispersion_mag})-(\ref{eq:heat_flux_mag}) is plotted on Fig. \ref{fig:heat_flux_mag} for different values of $\mathcal{P}$. When the magnetic field amplitude is higher, $\mathcal{P}$ is higher, and the overall heat-flux diminishes. The reduced wavenumber $z_m$ for which the flux is maximal is then higher. Directly from Eq.(\ref{eq:ratio_k}), the convective wavenumber is higher, which means the convective length scale is lower.

\begin{figure}[h!]
    \centering
    \includegraphics[width=\linewidth]{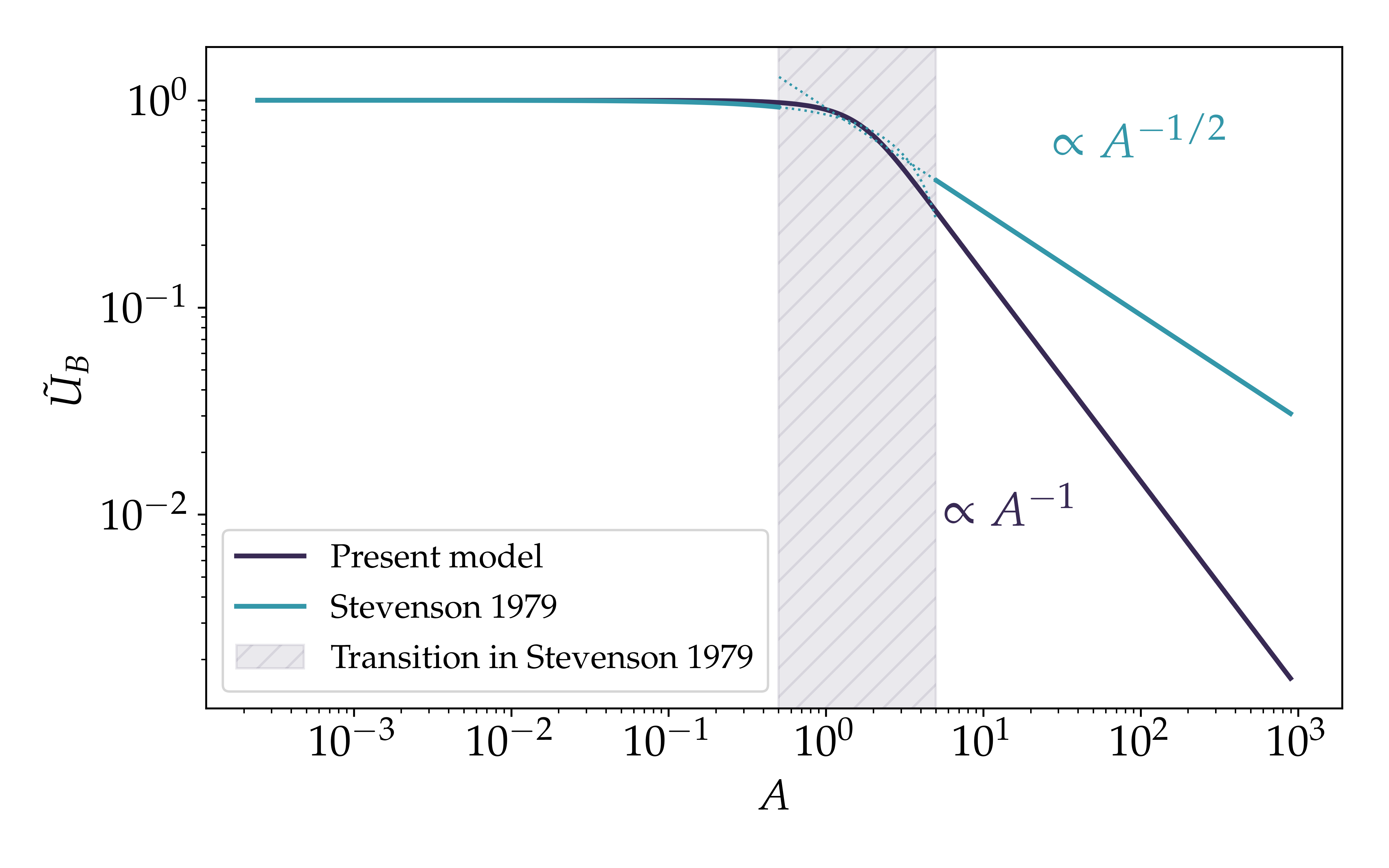}
    \includegraphics[width=\linewidth]{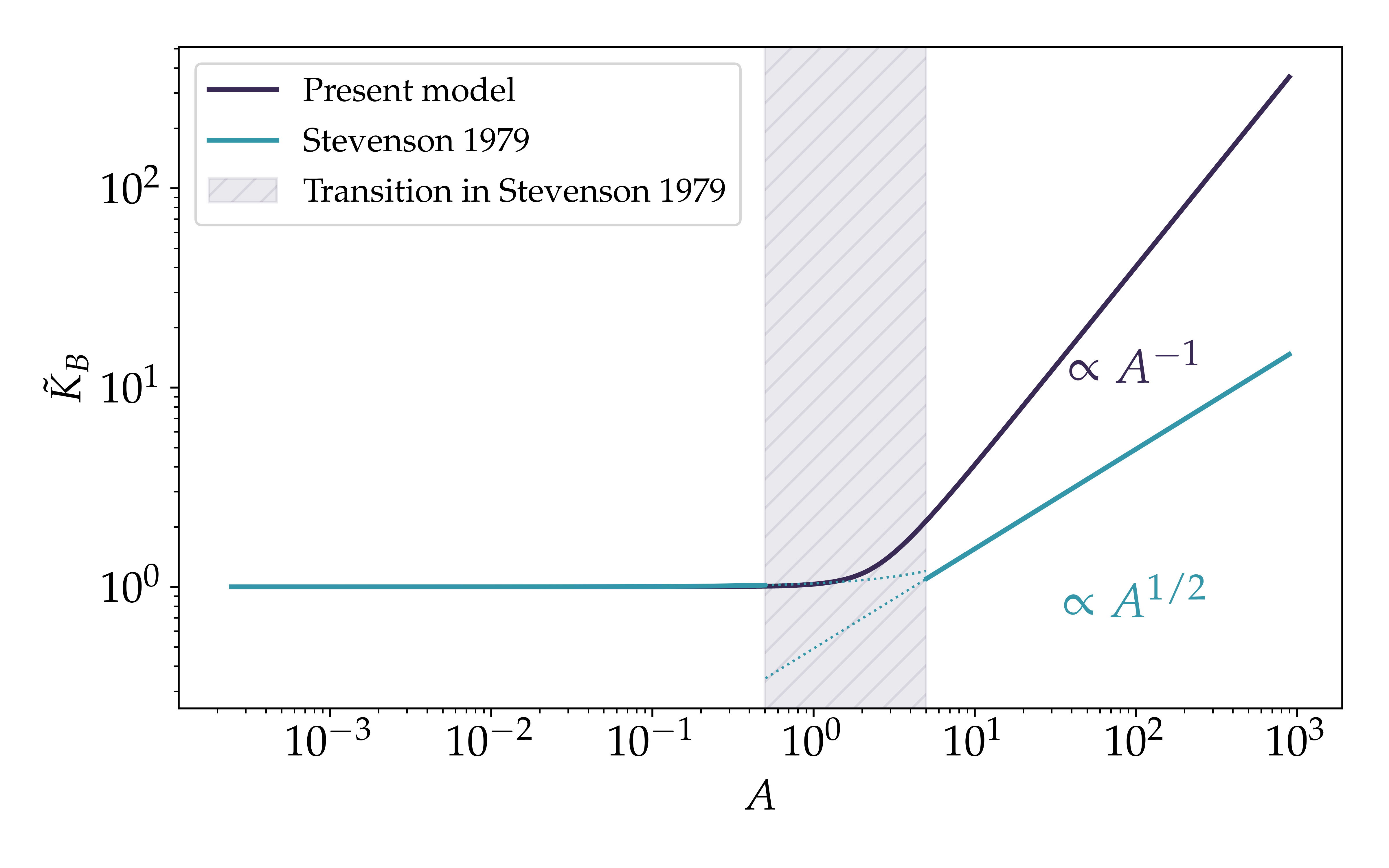}
    \includegraphics[width=\linewidth]{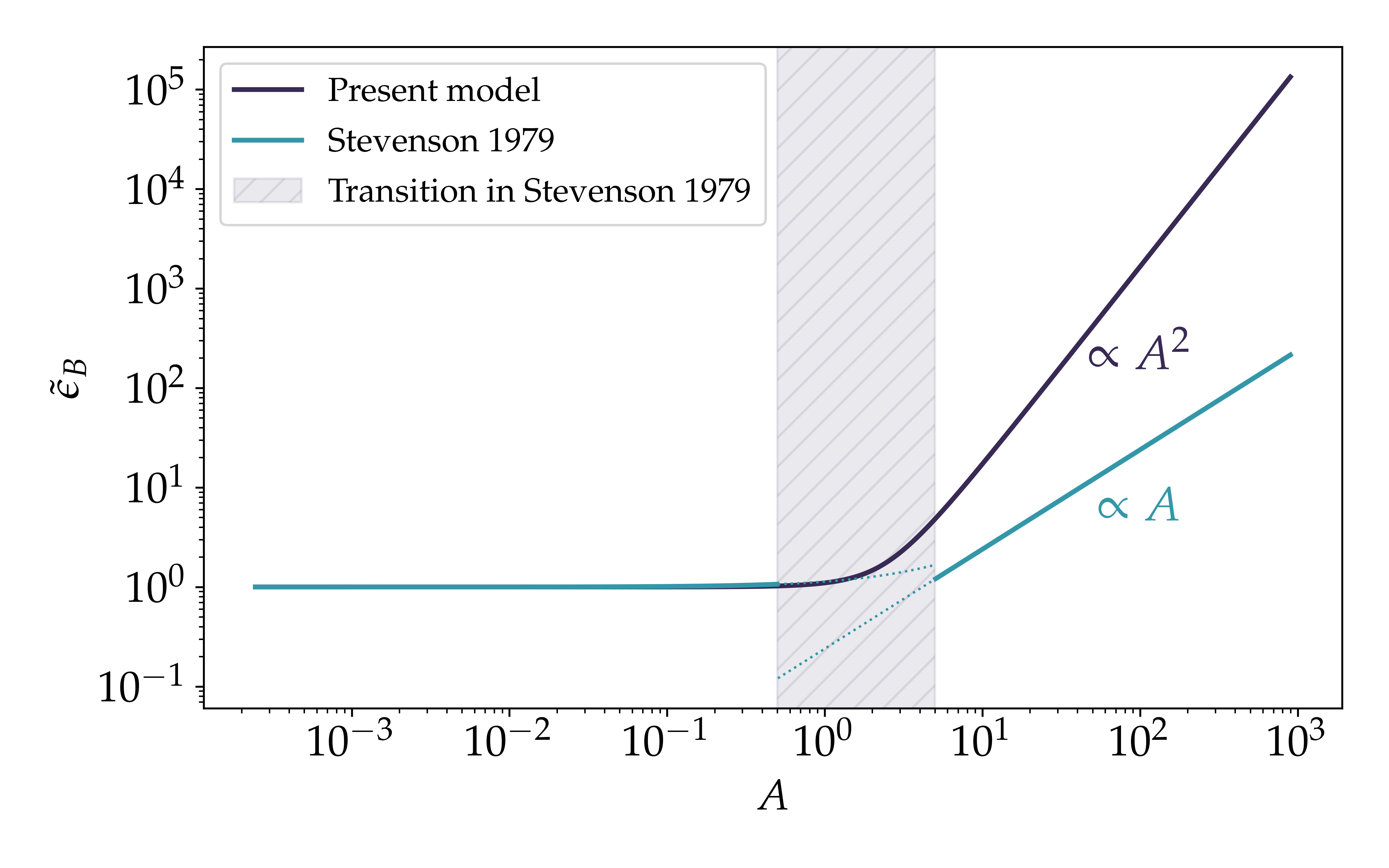}
    \caption{M-MLT results from \cite{stevenson_turbulent_1979} and the present work. Top: Convective velocity modulation with magnetism $\tilde{U}_B$ as a function of the inverse Alfvén number. Center: Convective wavenumber modulation with magnetism $\tilde{K}_B$ as a function of the inverse Alfvén number. Bottom: Superadiabaticity modulation with magnetism $\tilde{\epsilon}_B$ as a function of the inverse Alfvén number.}
    \label{fig:mmlt}
\end{figure}

The results for the modulation of the convective wavenumber,  velocity and superadiabaticity are shown in Fig. \ref{fig:mmlt} and compared with \cite{stevenson_turbulent_1979} asymptotic expressions (see Tab. \ref{tab:stevenson_mag}) for the low-$A$ and the high-$A$ regimes. As \cite{stevenson_turbulent_1979} does not account for the transition around $A \sim 1$, we represented this region with a hatched background. The tendencies for the magnetised case are the same for our study and the one from \cite{stevenson_turbulent_1979}: the higher the magnetic field strength, the higher the Alfvén number and the lower the convective velocity and the higher the convective wavenumber. As detailed in Sec. \ref{sec:studies_magnetic}, this is coherent with the general idea that an imposed magnetic field tends to inhibit convection through the Lorentz force. However, we predict a sharper transition from the low-$A$ to the high-$A$ case compared to \cite{stevenson_turbulent_1979}: the velocity modulation is nearly constant until $A \sim 10^{-1}$, and for high values of $A$ we find the convective velocity modulation scales as $\tilde{U}_B \sim A^{-1}$.

In this approach, we then bridge the gap between the low-$A$ case and the high-$A$ case. This region is paramount for the convective dynamics. As shown in numerical simulations of the dynamo process, equipartition of the magnetic energy with the turbulent kinetic energy seems a reasonable estimate in many contexts ranging from stars to planets \citep[see e.g.][]{christensen_scaling_2006, christensen_energy_2009, brun_magnetism_2017}. Such values for the magnetic fields also agree relatively well with observations \citep{christensen_energy_2009}, although some brown dwarfs might exhibit higher values \citep{kao_auroral_2015}. In addition, some numerical simulations find fields with values higher than equipartition, which are said to be in "super-equipartition" \citep{augustson_magnetic_2016}. A magnetic field in equipartition results in an Alfvén number $A = 1$, so the transition around this number is key to accurately modelling magnetised convection zones.
Moreover, we find the convective velocity (resp. convective wavenumber) decreases (resp. increases) more rapidly when the inverse Alfvén number increases compared to \cite{stevenson_turbulent_1979} study. In the highly magnetized case, where $A \gg 1$, we find the following power laws: 
\begin{equation}
    \tilde{U}(A) \approx 1.218 \cdot A^{-1.000}, 
\end{equation}
\begin{equation}
    \tilde{K}(A) \approx 0.484 \cdot A^{1.000},
\end{equation}
\begin{equation}
    \tilde{\epsilon}(A) \approx 0.240 \cdot A^{2.000}. 
\end{equation}

\section{Rotating and magnetised convection}
\label{sec:rmmlt}
\subsection{Previous studies of rotating and magnetised convection}

As discussed earlier, rotation and a magnetic field, when acting individually, tend to suppress convection, particularly the mode that transports the most heat. However, when these two factors act together, the onset of convection exhibits more complex behavior. The interaction between thermal convection, rotation, and a magnetic field can be categorized into two main types based on the magnetic Reynolds number (\(\mathcal{R}e_m\)), which is the ratio of magnetic advection to magnetic diffusion. When \(\mathcal{R}e_m \ll 1\), the self-induced magnetic field is weak compared to the imposed magnetic field. Linear theory suggests that a magnetic field can mitigate the Taylor-Proudman constraint, which restricts variations along the rotation axis, thereby enhancing convection \citep[e.g.,][]{chandrasekhar_hydrodynamic_1961, eltayeb_hydromagnetic_1977, simitev_onset_2021}. Experimental studies have confirmed these theoretical predictions for the onset of rotating magnetoconvection \citep[e.g.,][]{nakagawa_experiment_1955, aurnou_connections_2020, king_magnetostrophic_2015}. These studies demonstrate that the Lorentz force can facilitate convection compared to non-magnetized rotating convection. Specifically, in the magnetostrophic regime, where the Coriolis force and the Lorentz force are approximately balanced, the critical Rayleigh number for rotating magnetoconvection can be lower than in the non-magnetized case, making convection easier to initiate. Additionally, two distinct stationary modes with significantly different characteristic length scales coexist \citep[e.g.,][]{chandrasekhar_hydrodynamic_1961, eltayeb_hydromagnetic_1977, aujogue_onset_2015} in this regime. In contrast, when \(\mathcal{R}e_m \gg 1\), the system can generate and sustain a magnetic field through dynamo action. Convection with dynamo has been extensively studied using Direct Numerical Simulations in geophysical and astrophysical contexts \citep[see e.g.,][]{glatzmaiers_three-dimensional_1995, guervilly_generation_2015, brun_magnetism_2017, kapyla_simulations_2023}. \cite{stevenson_turbulent_1979} extended his Mixing-Length Theory model to address rotating and magnetized convection in specific geometric configurations and under high-rotation and high-magnetic-field limits. To our knowledge, these theoretical prescriptions have not been compared with numerical simulations nor implemented in the context of stellar physics.
\begin{figure*}
    \centering
    \includegraphics[width=0.49\linewidth]{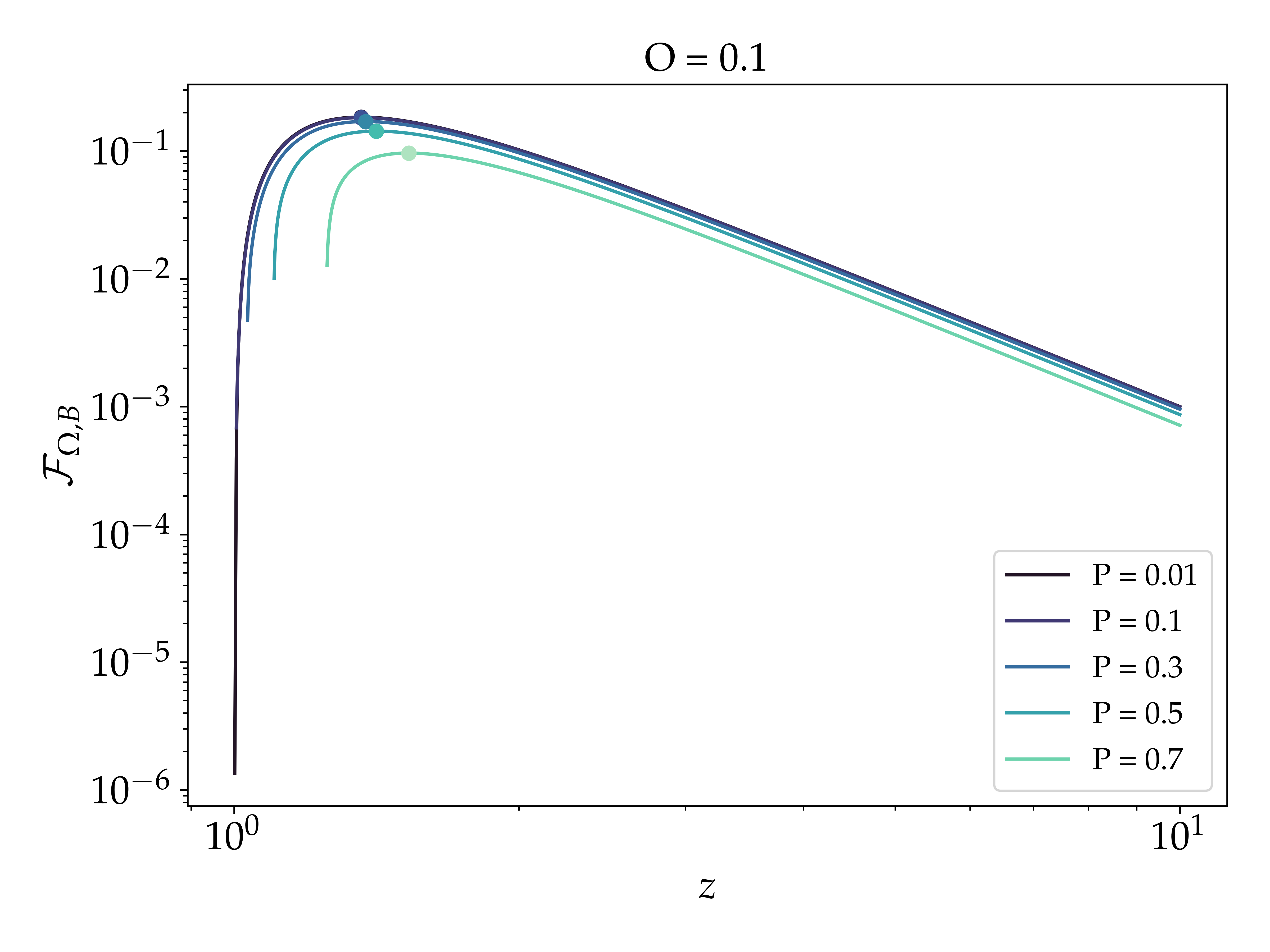}
    \includegraphics[width=0.49\linewidth]{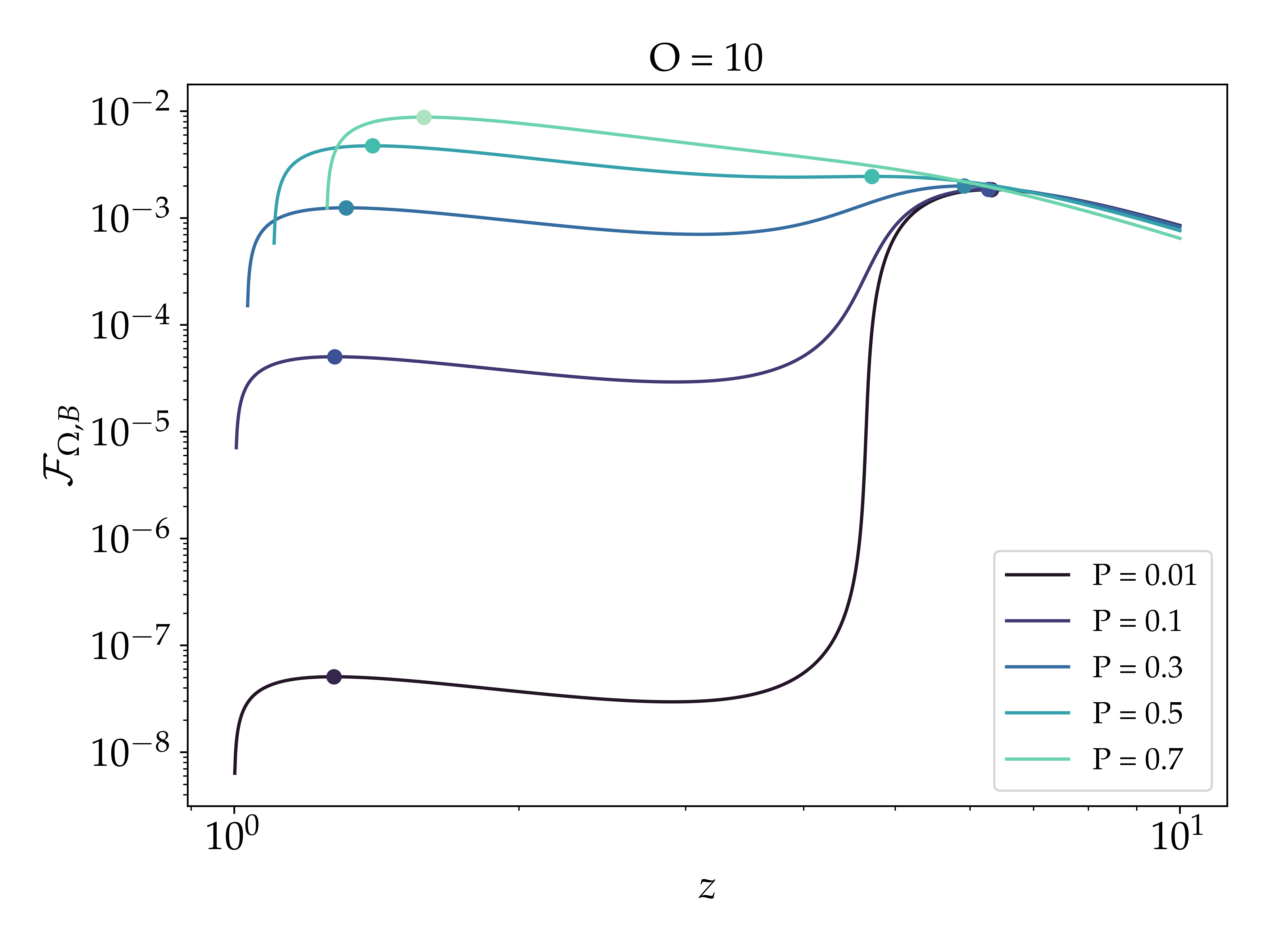}
    \caption{Convective heat-flux for rotating and magnetised convection, for the rotation parameter $\mathcal{O} = 0.1$ (left panel) and   $\mathcal{O} = 10$ (right panel), and several values for the magnetic parameter $\mathcal{P}$, as a function of the reduced wavenumber $z$. The dots denote the local maxima of the heat flux.}
    \label{fig:flux_omega_b}
\end{figure*}

\begin{figure*}[h!]
    \centering
    \includegraphics[width=0.48\linewidth]{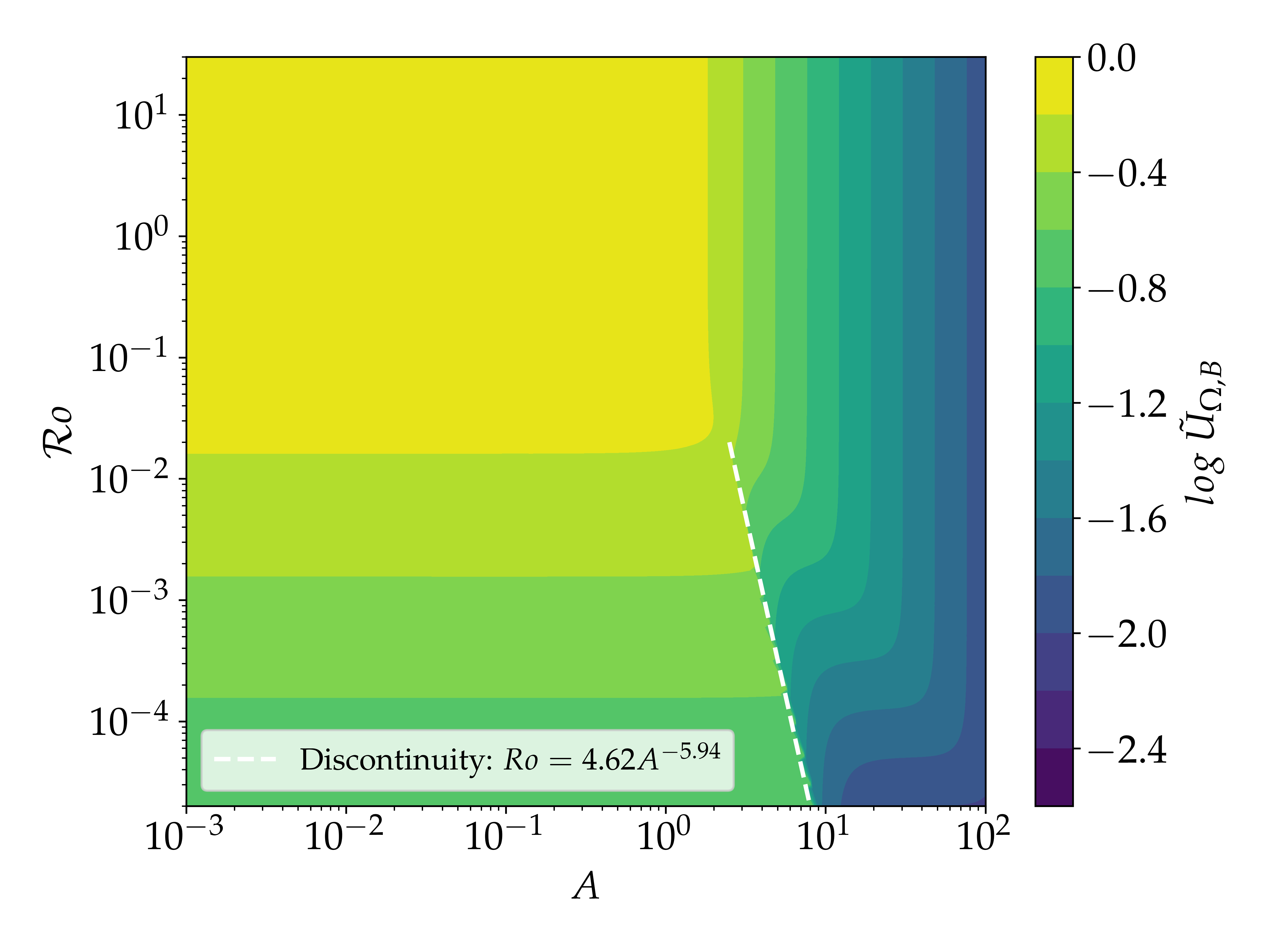}
    \includegraphics[width=0.48\linewidth]{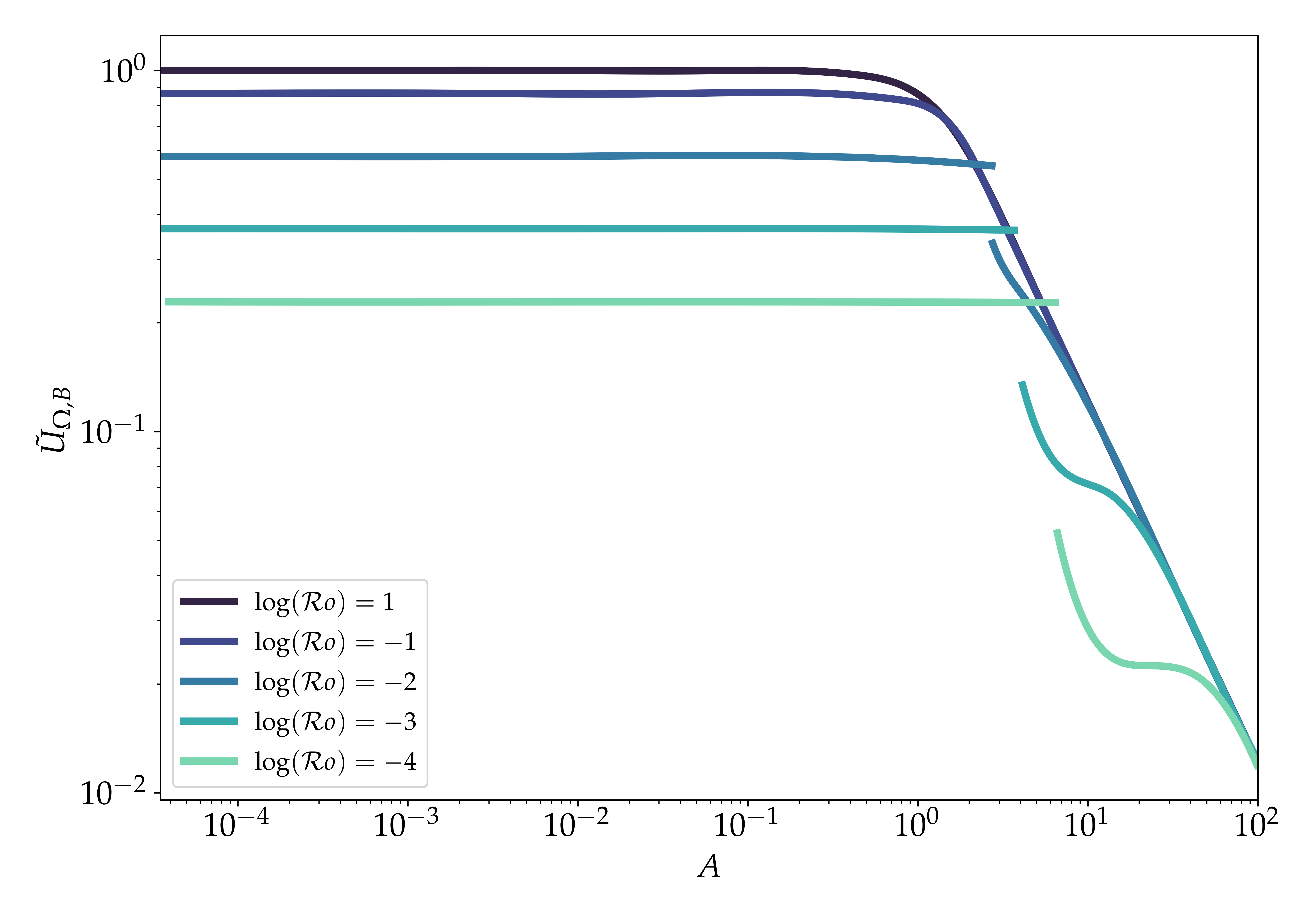}
    \includegraphics[width=0.48\linewidth]{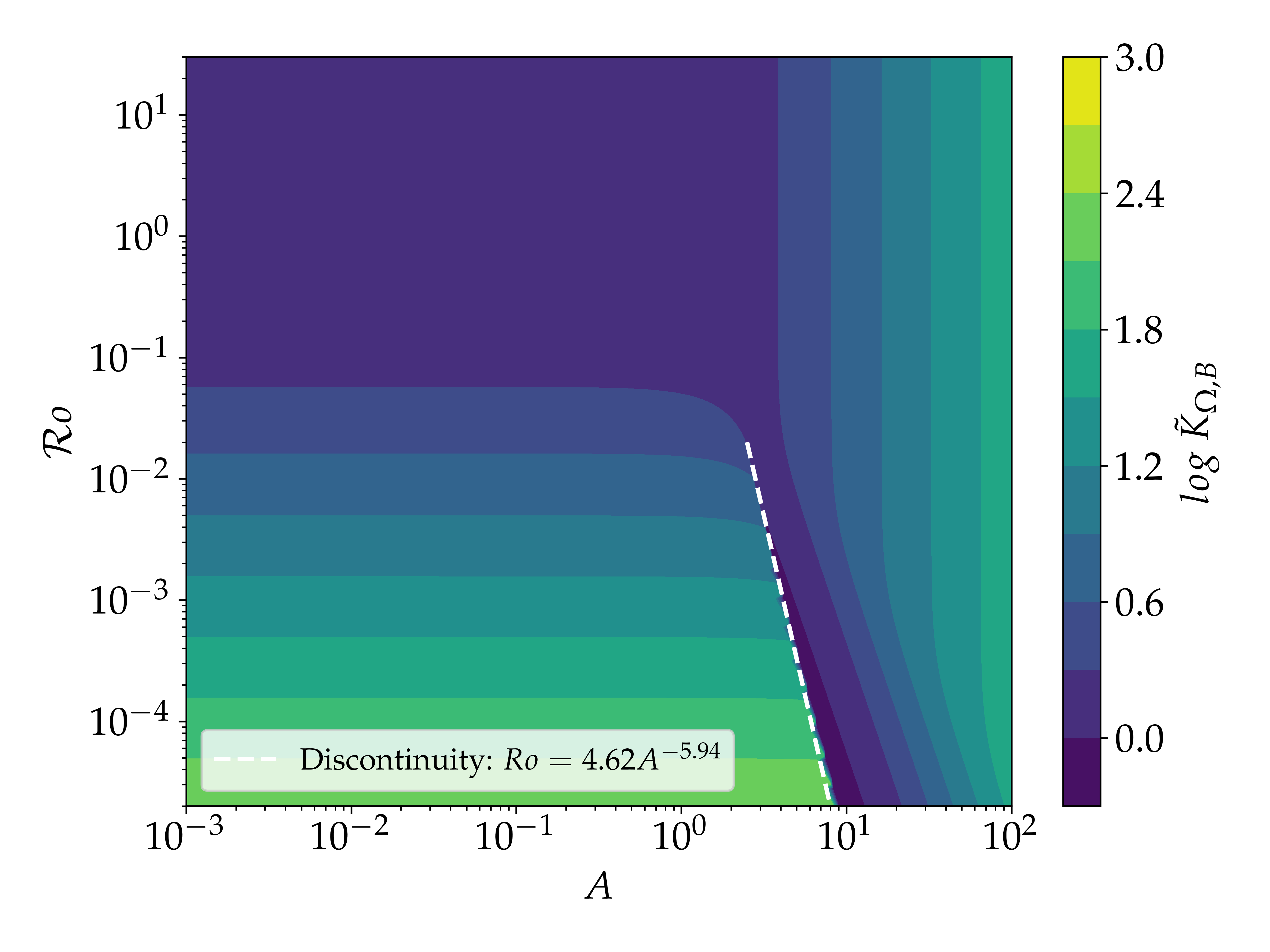}
    \includegraphics[width=0.48\linewidth]{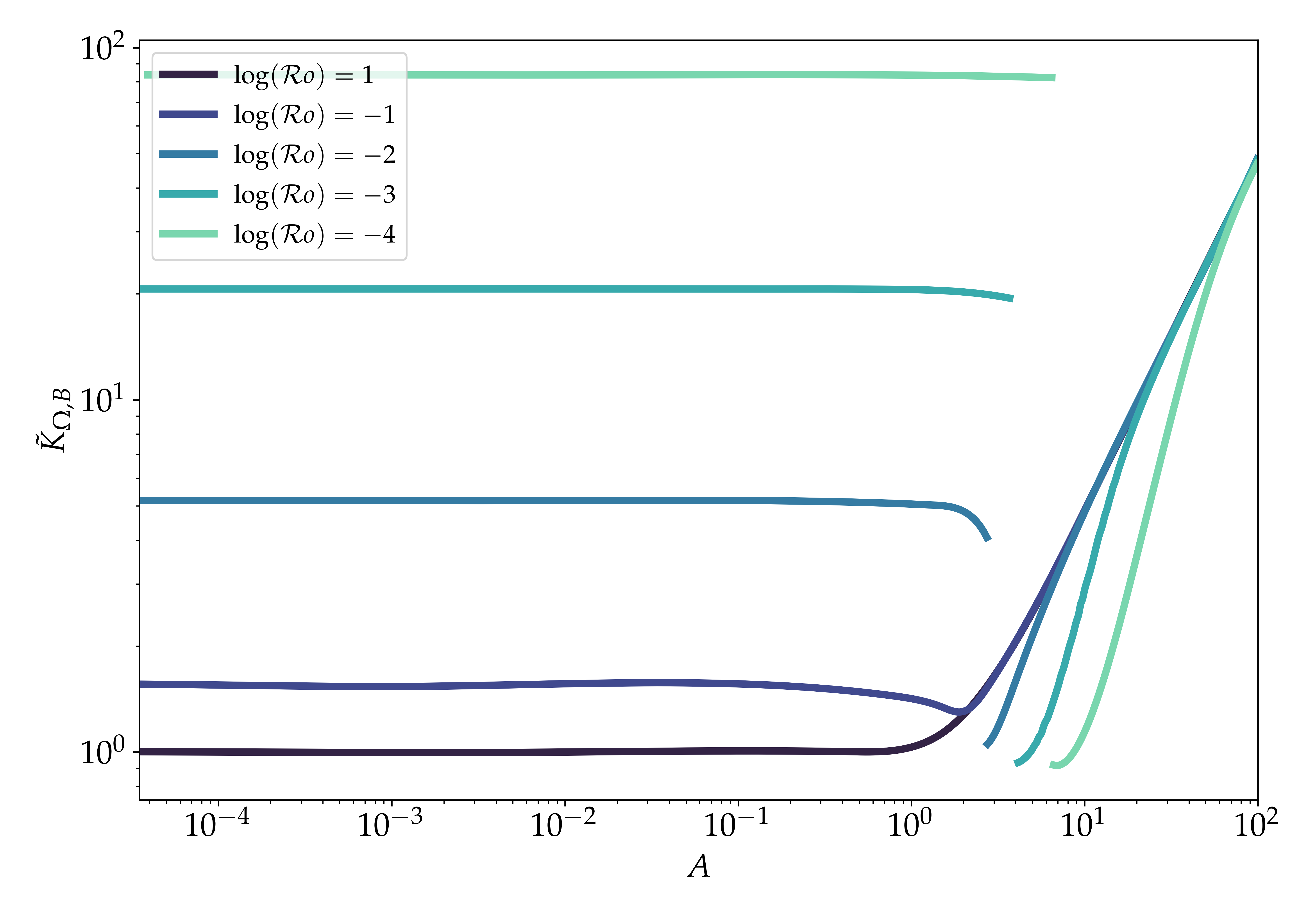}
    \includegraphics[width=0.48\linewidth]{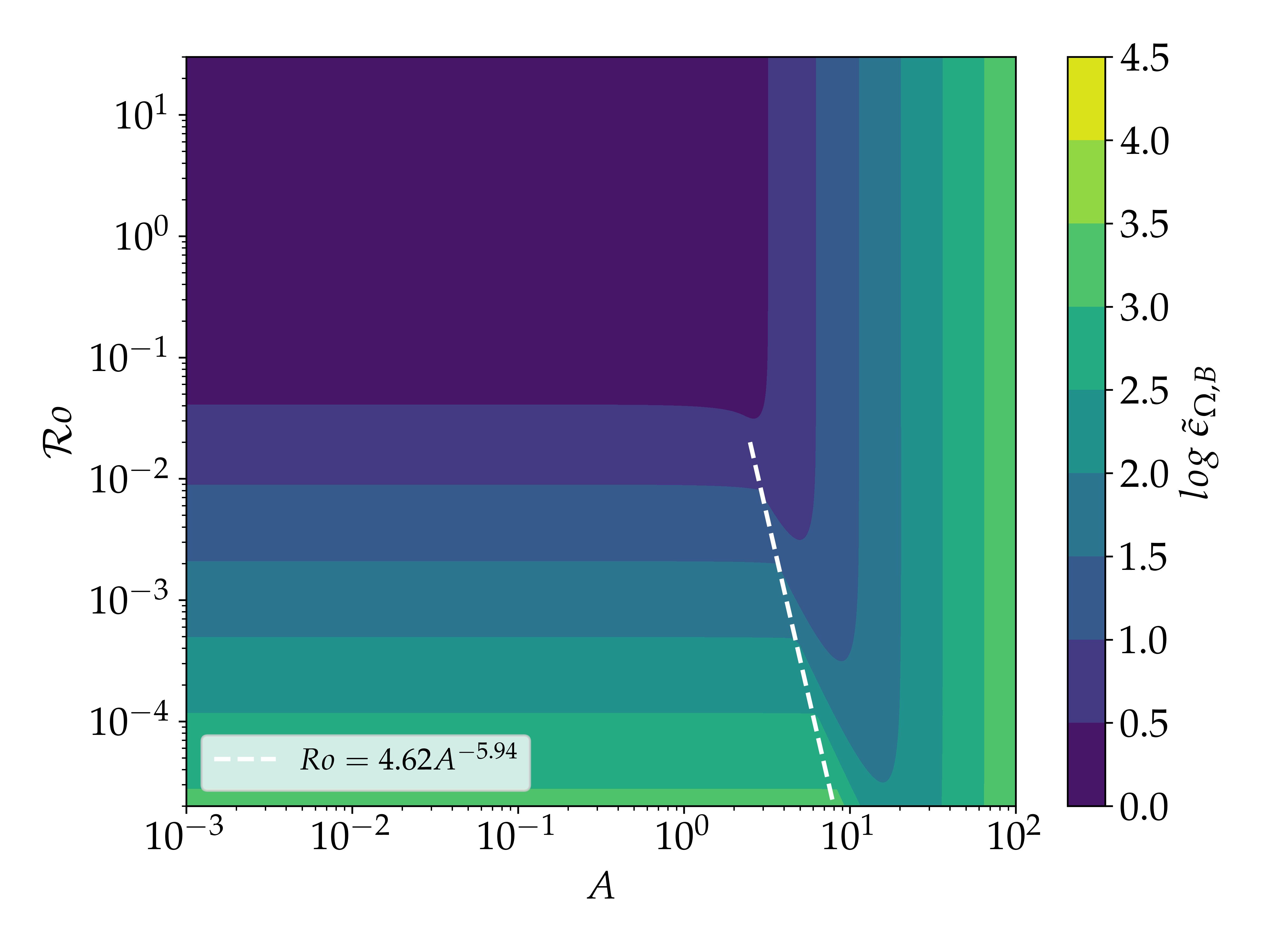}
    \includegraphics[width=0.48\linewidth]{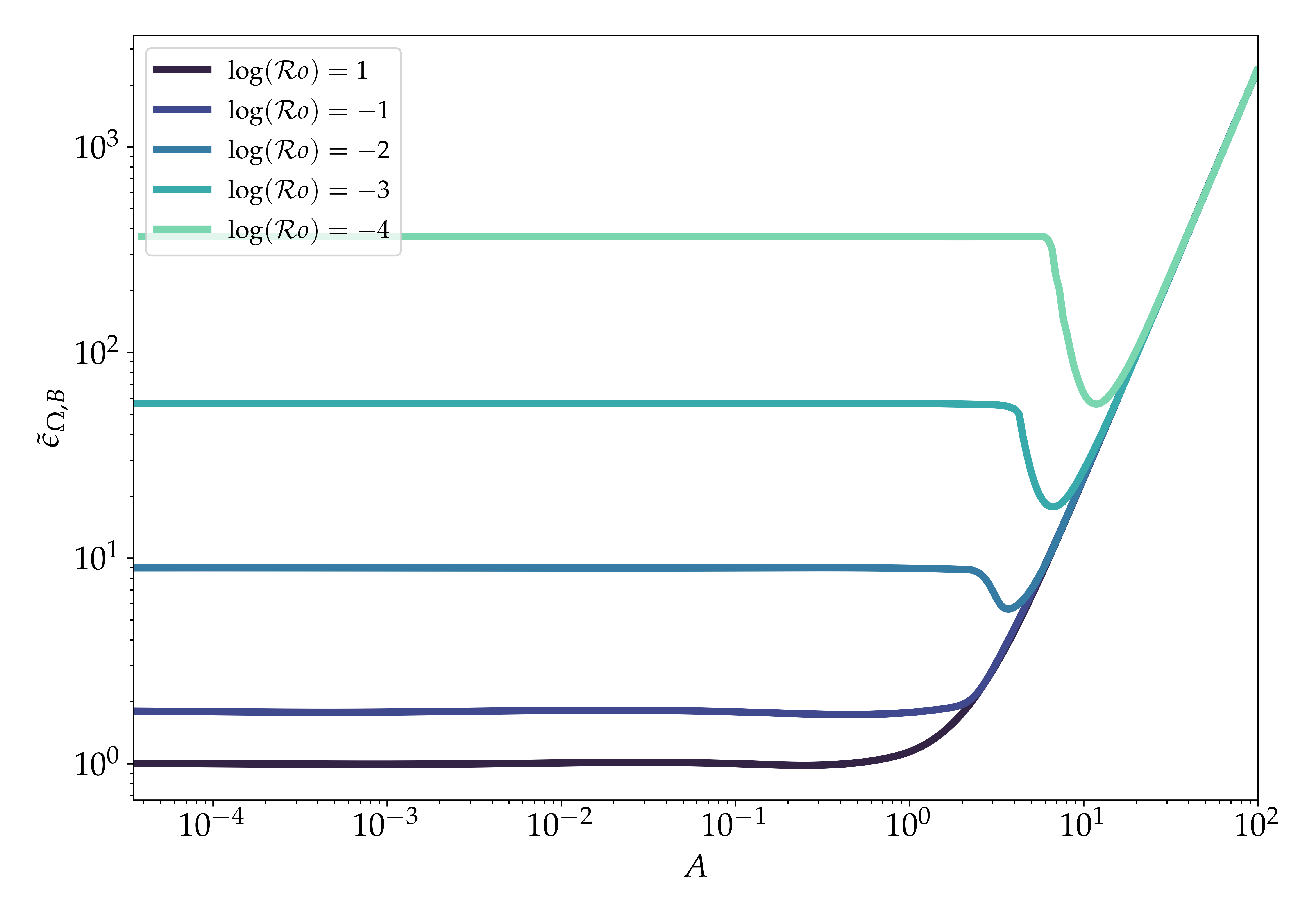}
    \caption{Top: Modulation of the convective velocity $\tilde{U}_{\Omega,B}$ Center: Modulation of the convective wavenumber $\tilde{K}_{\Omega,B}$ Bottom: Modulation of the superabadiaticity $\tilde{\epsilon}_{\Omega,B}$}
    \label{fig:vk_omega_b}
\end{figure*}

\subsection{Rotating and Magnetised Mixing-Length Theory (RM-MLT)}
In this section, we extend the Mixing-Length Theory model based on the heat-flux maximisation principle \citep{malkus_heat_1954, stevenson_turbulent_1979} to a rotating and magnetised framework. We employ the same method as in Sections \ref{sec:rmlt} and \ref{sec:mmlt}, taking both a non-zero rotation and a non-zero magnetic field: $\mathcal{O} \neq 0$ and $\mathcal{P} \neq 0$. 
As in the previous sections, we use  Eqs. (\ref{eq:dispersion_relation_normalised}), (\ref{eq:heat_flux_equal}) and (\ref{eq:heat_flux}) to maximise the heat-flux and determine the values of $z_m, \hat{s}_m, q_m$ such that the heat-flux is maximum. Eqs. (\ref{eq:rossby}) and (\ref{eq:alfven}) allow us to compute the corresponding Rossby $\mathcal{R}o$ and inverse Alfvén number $A$, while we use Eqs. (\ref{eq:ratio_k}), (\ref{eq:ratio_v}) and (\ref{eq:ratio_eps}) to calculate the convective velocity, wavenumber and superadiabaticity modulations, respectively, when compared to the non-rotating and non-magnetised case. 
\par The corresponding heat-flux is plotted in Fig. \ref{fig:flux_omega_b} for different values of $\mathcal{O}$ and $\mathcal{P}$. In the weak rotation case $\mathcal{O} = 0.1$, we recover the same tendency as in Fig. \ref{fig:heat_flux_mag}: for higher values of $\mathcal{P}$ corresponding to more magnetised cases, the heat flux is lower and the maximising wavenumber $z_m$ is higher (see Fig. \ref{fig:flux_omega_b}, left). However, for some given values of $\mathcal{O}$ and $\mathcal{P}$, we observe that two local maxima exist, suggesting that two linear modes are favoured (see Fig. \ref{fig:flux_omega_b}, right). This two-local-maxima situation may be linked to the magnetostrophic regime. In their linear stability analysis, \cite{horn_elbert_2022} find that two modes are dominant in this regime due to two local minima for the Rayleigh number. Assuming dominant modes maximise the convective heat-flux \citep{malkus_heat_1954}, the presence of two local maxima for the heat-flux implies that two convective modes govern. 
The present model of RM-MLT aims at providing the wavenumber and velocity modulation to be implemented in various astrophysical situations, requiring a single solution for the dominant mode's wavenumber. To align with \cite{malkus_heat_1954} theory, we always chose the global maximum for the heat-flux in this RM-MLT. However, one must keep in mind that several modes with different associated wavenumbers might dominate such a convective system. In this way, we notice a discontinuity: for values around $\mathcal{O} = 10$ and $\mathcal{P} = 0.5$, the local maximum around $z = 1$ becomes the global maximum for the heat-flux. This causes the dominant mode to shift abruptly from one value to another. 
\par We plot the results for the convective velocity modulation $\tilde{U}$, wavenumber modulation $\tilde{K}$ and superadiabaticity modulation $\tilde{\epsilon}$ in Fig. \ref{fig:vk_omega_b}. In all figures, we notice the discontinuity around $\mathcal{R}o \sim 2.62 A^{-5.94}$. In the strong magnetic field (resp. high rotation) limit, we recover the asymptotic tendencies discussed in the last sections \ref{sec:rmlt} and \ref{sec:mmlt}. In the $A \gg 1$ limit (resp. $\mathcal{R}o \ll 1$ limit), the convective velocity diminishes, and the wavenumber and superadiabaticity increase, indicating less efficient convection. However, this diminution in the convective velocity is sharper when going from the left to the right of the discontinuity, for $A \sim 1-7$ and $\mathcal{R}o \lesssim 5 \cdot 10^{-2}$. For the wavenumber modulation $\tilde{K}_{\Omega. B}$, the same trend emerges: overall, both rotation and magnetic field tend to increase the convective wavenumber in the $A \gg 1$ and $ \mathcal{R}o \ll 1$ regimes, respectively. However, when going from the left to the right of the discontinuity, the wavenumber sharply diminishes, indicating a larger convective length scale. The sharp decline in $\tilde{K}$, reaching up to two orders of magnitude for the lowest Rossby numbers, leads to a corresponding reduction in the convective velocity modulation $\tilde{U}$ to satisfy the constraints imposed by the conservation equation Eq. (\ref{eq:conservation_flux}) in the $A \sim 1$ regime. The predicted enlargment of the convective lengthscale in this regime has been reported experimentally by \cite{nakagawa_experiments_1957} in his set-up to study rotating magnetoconvection in liquid mercury. He has observed that the convective horizontal lengthscale increases in the magnetostrophic regime where the Coriolis and the Lorentz forces are approximately equal. It is important to note that such an increase has not been observed in recent numerical simulations by \cite{horn_elbert_2025}, which are the numerical counterparts of \cite{nakagawa_experiments_1957} experiments. Such a disagreement, likely caused by different boundary conditions, deserves further investigations in a near future.
In contrast, the superadiabaticity modulation $\tilde{\epsilon}_{\Omega. B}$ presents no discontinuity. Overall, a high rotation or a strong magnetic field tends to increase the superadiabaticity, resulting in less efficient convection. However, in the regime near the discontinuity for $A \sim 1-9$ and $\mathcal{R}o \lesssim 10^{-1}$, for a fixed Rossby number, increasing the magnetic field results in a lower superadiabaticity, indicating that the convective transport is more efficient. This aligns with the idea that the combination of magnetic field and rotation might facilitate convection in some cases \citep[e.g.][]{chandrasekhar_hydrodynamic_1961, horn_elbert_2022}. In this regime, rotation and magnetic field together produce a more efficient convection than the case with only rotation or magnetic field.

\subsection{Dynamo scaling laws}

In stars and planets, the magnetic field is generated by turbulent dynamo and is not independent of the rotation period. Scaling laws are often used to assess the magnetic field generated by dynamo effect, resulting from simple balance conditions of force or energy. They are then calibrated using observations or 2D or 3D simulations of low-mass stars and planets \citep[e.g.][]{christensen_energy_2009, augustson_rossby_2019}. \\

In the equipartition dynamo, the kinetic energy is converted into magnetic energy without losses. This equipartition regime is widely used to provide a reasonable order-of-magnitude estimate for the large-scale magnetic field in active regions at the surface of stars and in sunspots within their convective envelopes \citep[e.g.][]{donati_magnetic_2009, brun_magnetism_2017}. In this regime, the inverse Alfvén number is equal to unity \footnote{Note that sometimes an $\alpha$ parameter is introduced to mimic energy losses by ohmic heating. }: 
\begin{equation}
    A_{\rm eq} \sim 1.
\end{equation}

The buoyancy dynamo regime is referred to as MAC balance (Magneto-Archimedean-Coriolis) in geophysics \citep[e.g.][]{braginskiy_magnetic_1967}. In this regime, the Coriolis, buoyancy and Lorentz forces are considered to be of comparable magnitude. The regime is of particular interest for the geodynamo \citep[e.g.][]{christensen_energy_2009} or for low-mass stars \citep{augustson_model_2019}. For the boyancy dynamo, the inverse Alfvén number scales like \citep[see e.g. the appendix in ][]{astoul_does_2019}: 
\begin{equation}
    A_{\rm buoy} \sim \mathcal{R}o^{-1/4}.
\end{equation}

In the magnetostrophic dynamo, the Lorentz force balances the Coriolis force in the Navier-Stokes equation. In this regime, we have \citep[see e.g. the appendix in ][]{astoul_does_2019}: 
\begin{equation}
    A_{\rm mag} \sim \mathcal{R}o^{-1/2}.
\end{equation}

\begin{figure}
\centering
\includegraphics[width=1\linewidth]{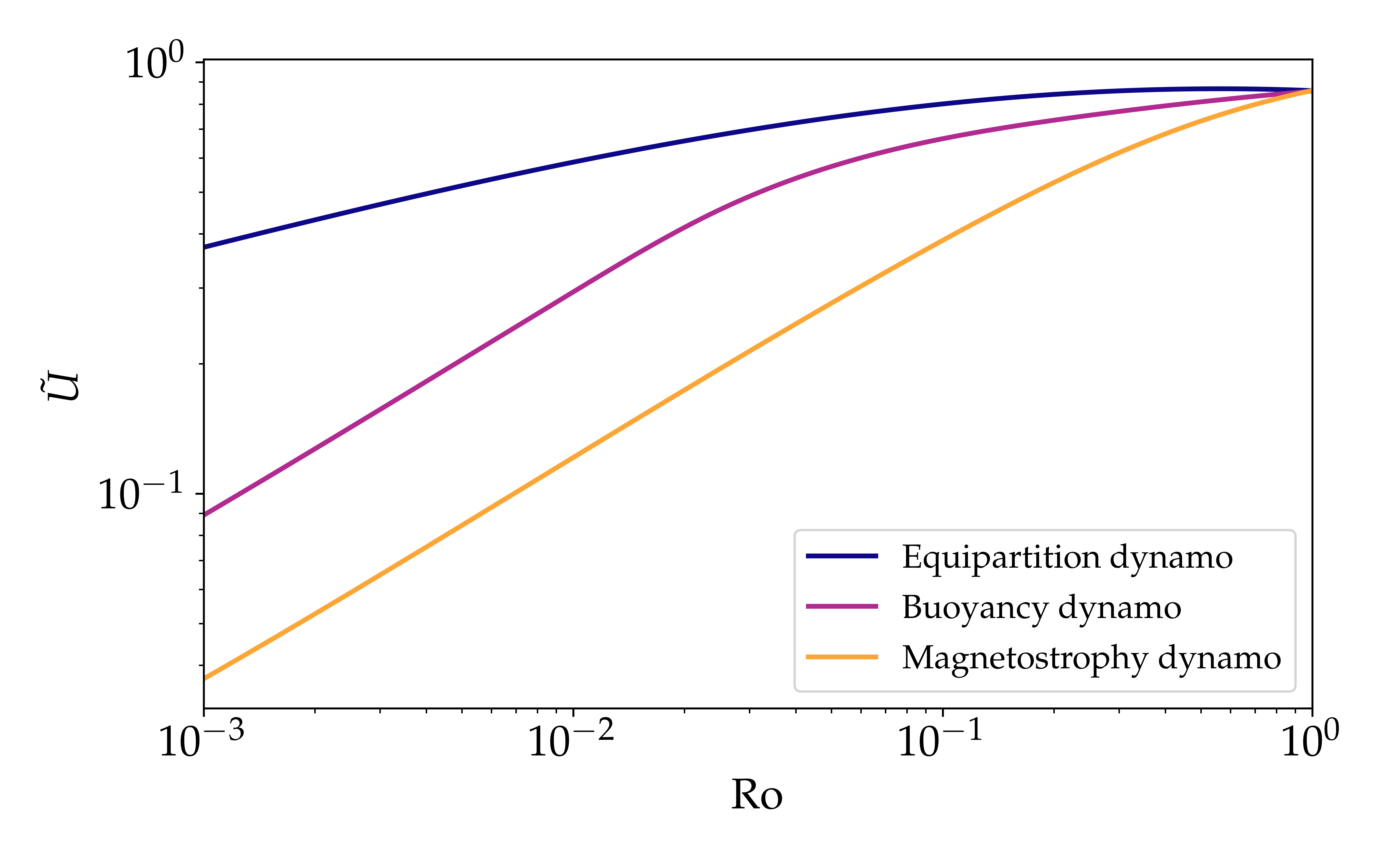}
\includegraphics[width=1\linewidth]{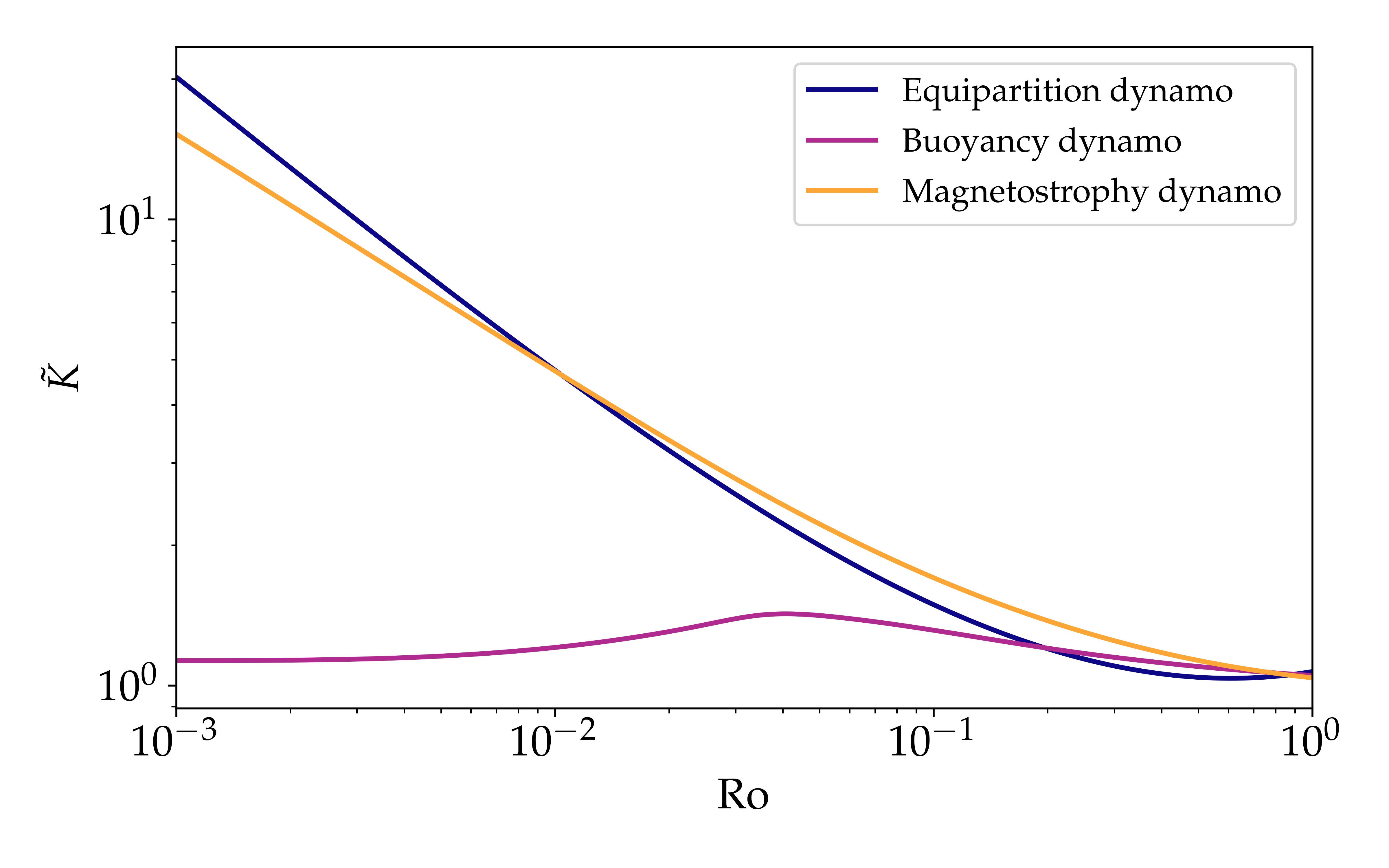}
\includegraphics[width=1\linewidth]{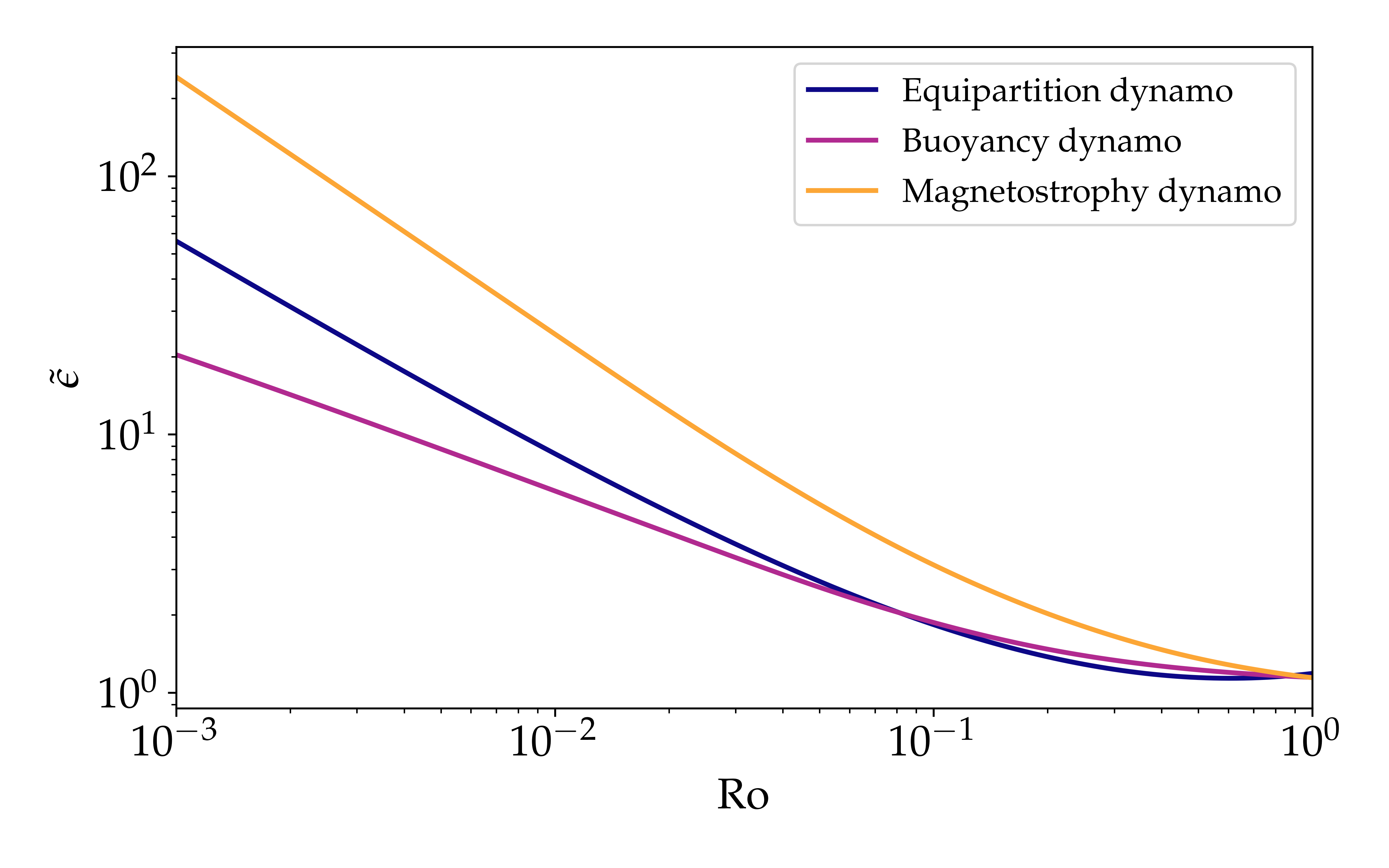}
\caption{Top: Modulation of the convective velocity $\tilde{U}$ in the equipartition, buoyancy and magnetostrophy dynamo regimes. Center: Modulation of the convective wavenumber $\tilde{K}$. Bottom: Modulation of the superadiabaticity $\tilde{\epsilon}$.}

\label{fig:scaling_dyanmo}
\end{figure}

Using these scaling laws to assess the inverse Alvén number in the different dynamo regimes, we plot the convective velocity (resp. wavenumber, superadiabaticity) modulation in Fig. \ref{fig:scaling_dyanmo} as a function of the Rossby number. At the top of Fig. \ref{fig:scaling_dyanmo}, the convective velocity decreases less in the equipartition regime than in the buoyancy and magnetostrophy ones. This is in line with the idea that magnetostrophy provides an upper bound for the magnetic field strength, while equipartition gives a lower estimate \citep[see e.g. the work of ][]{astoul_does_2019}. In magnetostrophy, the convective velocity in then further inhibited by the magnetic field. However, a different trend emerges in the modulation of the convective wavenumber: for both the equipartition and magnetostrophy dynamo regimes, the convective wavenumber increases when the Rossby number increases, while in the buoyancy regime the convective wavenumber modulation stays around 1. This is due to the presence of the discontinuity in Fig. \ref{fig:vk_omega_b}: the buoyancy dynamo regime lies to the right of this discontinuity, where the convective wavenumber modulation $\tilde{K}$ remains low. Finally, the superadiabaticity increases when the Rossby number decreases. The strongest increase is observed in the magnetostrophy dynamo, followed by the equipartition and then the buoyancy ones. This result suggests that the buoyancy dynamo would lead to more efficient convection than the other regimes, which contrasts with the theoretical, experimental and numerical works that identify the magnetostrophic balance as optimally efficient \citep[e.g.][]{king_magnetostrophic_2015}. This apparent discrepancy may arise from the fact that the scaling laws we use here are applied in a simplified framework. They serve as first-order estimates of the inverse Alfvén number based on force balance arguments. While they offer valuable insight into the general trends associated with different possible dynamo regimes, they do not capture the full complexity of nonlinear magnetoconvection, nor the feedbacks between convection, magnetic fields and rotation. It would be valuable to further refine these scaling relations and their impacts on the Rotating and Magnetised Mixing-Length Theory with direct numerical simulations.

\section{Conclusion}
\label{sec:conclusion}
\subsection{Summary and discussion}
We present in this work an extension of the standard Mixing-Length Theory for rotating convection (R-MLT), magnetised convection (M-MLT) and rotating and magnetised convection (RM-MLT). Building upon the heat-flux maximisation principle \citep{malkus_heat_1954}, we extend the previous work by \cite{stevenson_turbulent_1979} and \cite{augustson_model_2019} in Boussinesq convection, neglecting the effect of diffusivities ($\eta = \nu = \kappa = 0)$. This approach allows us to compute the modulation of the convective superabadiaticity, velocity and wavenumber, respectively, with rotation and/or an imposed magnetic field, as a function of the Rossby number and/or the inverse Alfvén number. Our study focuses on a simple geometrical setup in Cartesian coordinates where the rotation axis and/or the magnetic field are vertical. Our findings align qualitatively with previous theoretical \citep[e.g.][]{chandrasekhar_hydrodynamic_1961, augustson_model_2019} and numerical \citep[e.g.][]{barker_theory_2014,  hotta_breaking_2018,currie_convection_2020} works on rotating or magnetised convection: the higher the rotation or the magnetic field, the lower the convective velocity and the higher the convective wavenumber. The convective efficiency is then reduced by the action of rotation or magnetic fields. However, our model predicts a sharper diminution of the convective velocity (resp. convective lengthscale) in the magnetised case than in a similar study by \cite{stevenson_turbulent_1979}. Moreover, when both rotation and magnetic field are present together, we observe a regime with two dominant modes with distinct wavenumbers, which might be related to the magnetostrophy found in the literature \citep[e.g.][]{horn_elbert_2022}. To provide a simple model for the Mixing-Length Theory, we assume that the mode that makes the heat-flux maximum is dominant, excluding secondary modes from local heat flux maxima. This leads to a discontinuity in the wavenumber and convective velocity modulations. When going from the left to the right of this discontinuity in Fig. \ref{fig:vk_omega_b}, the convective velocity and the convective wavenumber decrease when the inverse Alfvén number increases. Moreover, in the region $A \sim 1-9$ and $\mathcal{R}o \lesssim 10^{-1}$ the superadiabaticity decreases with increasing magnetic field for a given Rossby number. This indicates that the convective transport is more efficient when increasing the magnetic field (resp. rotation) compared to the case with only rotation (resp. magnetic field) in this regime. This is in line with previous works regarding the linear stability of rotating magnetoconvection \citep[e.g.][]{chandrasekhar_hydrodynamic_1961}. Finally, as the magnetic field in stars and planets is generated by dynamo action, it is not independant in general from rotation. We used classical dynamo scaling laws to give estimates for the convective velocity (resp. convective wavenumber, superadiabaticity) modulation as a function of rotation exclusively. Our results show that the convective velocity decreases with decreasing Rossby number, this reduction being less pronounced in the equipartition regime than in the magnetostrophic and buoyancy regimes. In contrast, the convective wavenumber increases with decreasing Rossby number in both the equipartition and magnetostrophic dynamos, while it remains roughly constant in the buoyancy dynamo. Finally, the superadiabaticity increases with increasing rotation rate across all three regimes. \\

\subsection{Perspectives}
The perspectives of this work can be structured along three main axes. First, this model is a foundational step towards integrating rotation and magnetic fields in MLT for the modelling of stellar and planetary structure and evolution. Future improvements should include diffusive processes as in \cite{augustson_model_2019}. Upcoming refinements should also explore more complex geometries, such as an inclined magnetic field and rotation axis. In addition, the conjecture we base on from \cite{stevenson_turbulent_1979} focuses on the mode that maximises the heat flux. While this monomodal approach to convection is relevant for the MLT, which is widely used in stellar physics, the reality is more complex. Extending this theory to perform a mode by mode analysis of the heat-flux and the associated turbulent spectrum, similar to \cite{malkus_heat_1954}, or more recently to \cite{currie_convection_2020} for rotating convection, would provide a more comprehensive understanding. Secondly, the outputs from this heuristic theory of RM-MLT should be validated against Direct Numerical Simulations or experimental results, similar to previous studies of rotating convection \citep[e.g.][]{barker_theory_2014, currie_convection_2020}. Such comparison will assess the robustness of the RM-MLT prescriptions, and eventually refine its conclusions. 
\par Another perspective involves implementing the prescriptions from this model of RM-MLT into stellar structure and evolution numerical codes, exploring various stellar masses and evolutionnary stages. However, it is important to note that RM-MLT predicts the modification of wavenumeber and convective velocity for regions where convection is effectively unstable. The current model should be combined with a stability criterion for convection that includes both rotation and magnetic field (as in the critical magnetic field approach, \cite{macdonald_magnetic_2019}, or the critical Rayleigh number approach as in \cite{chandrasekhar_hydrodynamic_1961}) in order to correctly model rotating and/or magnetised convection in stars and planets.
\par Finally, convection plays a crucial role in many physical mechanisms occurring in stars and planets, making RM-MLT relevant for studying the impact of rotation and magnetic fields on such processes. For example, convective flows can penetrate into stably stratified layers, modifying the chemical composition, thermodynamical properties and mixing between convective unstable and stable regions \citep[e.g.][]{zahn_convective_1991}. Convective penetration and overshoot can also lead the stars to spend more time in the burning phases, as more fuel is mixed into the burning regions. Therefore, understanding how rotation and magnetism impact the penetration depth is key to correctly modelling stellar structure and evolution. The work by \cite{augustson_model_2019} in the case where only rotation is taken into account could be extended to take into account both rotation and magnetic fields. 
Moreover, some waves can be excited by turbulent convection. Rotation and magnetic field seem to have an impact on acoustic mode excitation, hindering mode detection in nearly $40\%$ of stars from \textit{Kepler} \citep{chaplin_predicting_2011, mathur_revisiting_2019}. Previous theoretical studies separately included the effects of magnetic fields \citep{bessila_stochastic_2024} and rotation \citep{bessila_impact_2024}. The present RM-MLT model should enable us to assess mode excitation with both rotation and magnetic fields. Internal gravity waves, which propagate in stably stratified stellar interiors and transport angular momentum and chemicals \citep[e.g.][]{press_radiative_1981, schatzman_transport_1993, zahn_angular_1997, kumar_angular_1997}, are excited stochastically by turbulent convection at the interface between the convective and the radiative zones. Thus, understanding the modifications of convection by rotation and magnetic fields is paramount for evaluating excitation of waves and the strength of the momentum transport and chemicals mixing they trigger \citep[e.g.][]{mathis_impact_2014, rogers_differential_2015,augustson_model_2020}. Finally, turbulent friction drives tidal flow dissipation in convective regions \citep[e.g.][]{zahn_marees_1966,zahn_tidal_1989,ogilvie_interaction_2012}. \cite{mathis_impact_2016} and \cite{devries_tidal_2023} incorporated the effects of rotation on tidal dissipation using \cite{stevenson_turbulent_1979} scaling laws. They showed that turbulent friction is significantly modified in a rapidly rotating framework. To go further, the present model of RM-MLT should be used to assess the impact of both rotation and magnetic fields. This would allow us to take into account the impact of the modification of convection in stars and planets because of the variations of their rotation and magnetic fields on the secular evolution of their structure, angular momentum and of their environment.

\FloatBarrier
\begin{acknowledgements}
L.B. and S. M. acknowledge support from the  European  Research Council  (ERC)  under the  Horizon  Europe program (Synergy  Grant agreement 101071505: 4D-STAR), from the CNES SOHO-GOLF and PLATO grants at CEA-DAp, and from PNPS (CNRS/INSU). While partially funded by the European Union, views and opinions expressed are however those of the author only and do not necessarily reflect those of the European Union or the European Research Council. Neither the European Union nor the granting authority can be held responsible for them.
\end{acknowledgements}


\bibliographystyle{aa}
\bibliography{references2-5_abbr}


\end{document}